\documentclass[superscriptaddress,12pt,tightenlines,aps,nofootinbib,pre]{revtex4-1}
\usepackage{color,epsfig,graphics,amssymb,amsmath,subeqnarray,graphicx,amsthm,subfigure,mathrsfs,bm,color,xcolor,soul}
\usepackage{newtxtext}
\usepackage{newtxmath}
\usepackage{caption}
\usepackage{natbib}
\usepackage{hyperref}
\hypersetup{
    colorlinks = true,
    urlcolor   = blue,
    citecolor  = black,
}
\newcommand{\RomanNumeralCaps}[1]

\def\bea{\begin{equation}}
\def\eea{\end{equation}}

\begin{document}
\title{Exact solutions for submerged von K\'arm\'an point vortex streets cotravelling with a wave on a linear shear current}
\author{Jack S. Keeler}
\affiliation{School of Mathematics, University of East Anglia,
Norwich, NR4 7TJ, UK}
\author{Darren G. Crowdy}

\affiliation{Department of Mathematics, Imperial College London,
180 Queen's Gate,
London, SW7 2AZ, UK}


\begin{abstract}
New exact solutions are presented to the problem of steadily-travelling water waves with
vorticity wherein a submerged von Kármán point vortex street cotravels with a wave on
a linear shear current. Surface tension and gravity are ignored. The work generalizes an
earlier study by Crowdy \& Nelson [Phys. Fluids, 22, 096601, (2010)] who found analytical
solutions for a single point vortex row cotravelling with a water wave in a linear shear
current. The main theoretical tool is the Schwarz function of the wave and the work builds
on a novel framework recently set out by Crowdy [J. Fluid Mech., 954, A47,
(2022)]. Conformal mapping theory is used to construct Schwarz functions with the requisite
properties and to parametrize the waveform. A two-parameter family of solutions
is found by solving a pair of nonlinear algebraic equations. This system of
equations has intriguing properties: indeed, it is degenerate, which radically reduces
the number of possible solutions, although the space of physically admissible equilibria is still found to be rich and diverse. Inline vortex streets, where
the two vortex rows are aligned vertically, there is generally a single physically admissible
solution. However, for staggered streets, where the two vortex rows are horizontally
offset, certain parameter regimes produce multiple solutions. An important outcome of the
work is that while only degenerate von Kármán point vortex streets can exist in an unbounded
simple shear current, a broad array of such equilibria are possible in a shear current beneath
a cotravelling wave on a free surface.

\end{abstract}

\maketitle

\section{Introduction}

In the study of two-dimensional water waves it is common to assume that the flow is irrotational 
and to study the effects of gravity, or capillarity, or both.
There is growing interest, however, in the theory of water waves with vorticity
where finite-amplitude steadily-travelling waves can exist even without either of these physical effects
\citep{Benjamin, Simmen, Peregrine, Pullin, VB1, VB2, VB3, Constantin1, Ehrn, Wahlen1, Wahlen2, 
Wahlen0, Hur2, Hur3, Hur1, Miles}.
A recent review article \citep{MilesReview} 
gives an overview of some of the literature on
water waves with vorticity.
When adding vorticity to the water wave problem, there is a choice on the form of the vorticity distribution
 and it has traditionally been taken to be uniform:
\cite{Tsao} 
and \cite{Benjamin}
performed early weakly nonlinear analyses of this case. 
\cite{Simmen} studied it numerically for gravity waves in deep water, work extended by
\cite{Peregrine} to the finite depth scenario. In the infinite depth case, one supposes that
at large distances from the interface the flow is 
 a linear shear current.
By now, much other numerical work has been done for constant-vorticity water waves using a variety of formulations
\citep{VB1, VB2, VB3, Hur2, Hur3, Hur1, Miles}.

Another vortex model that has been studied in the context of water waves is the point vortex.
Early work on the rigorous existence theory, when gravity is present but weak, and when
the vorticity is modelled as a point vortex, 
was carried out by \cite{Filippov} and \cite{Ter}.
\cite{Walsh1} have proved
the existence of steadily travelling two-dimensional capillary-gravity water waves 
with compactly supported vorticity, including the case where the vorticity is
in the form of point vortices.
\cite{Varholm} constructed solitary solutions for
gravity-capillary waves with a submerged point vortex.
\cite{LeSIAM} looked at solitary waves carrying a submerged finite dipole in deep water.

Recently, one of the authors \citep{Recent} has introduced a novel 
theoretical framework for understanding
water waves with uniform vorticity, in the absence of gravity or surface tension, 
and possibly also punctuated by rows of cotravelling point vortices.
The  mathematical tool used in this framework is the notion of a Schwarz function of a wave.
Earlier, \cite{Nelson} used Schwarz functions in the context of a water wave problem involving a linear shear current, and
the recent work of \cite{Recent} shows how that study fits into a broader framework. 

To explain the Schwarz function of a wave, consider first
a flat wave profile $y=0$, say, in a Cartesian $(x,y)$ plane. Using the complex variable $z=x+{\rm i}y$, clearly
\begin{equation}
\overline{z} = z, \qquad {\rm on}~y=0.
\label{flat}
\end{equation}
A key observation is that
the right hand side of (\ref{flat}) is an analytic function of $z$ having the feature that it can be analytically  continued off the line
$y=0$. 
For a more general wave profile, $\partial D$ say, 
given by an analytic curve that is periodic in the $x$ direction,
 the Schwarz function of $\partial D$ can be defined as the function
$S(z)$, analytic in a strip containing the wave profile, satisfying the conditions
\begin{equation}
 \overline{z} = S(z), \qquad {\rm on}~\partial D
\label{ffc0}
\end{equation}
with
\begin{equation}
S(z)  \to  z + {\rm i} \Lambda + {\cal O}(1/z), \qquad {\rm as}~y \to -\infty, \qquad \Lambda \in \mathbb{R}.
\label{ffc}
\end{equation}
The Schwarz function of the flat profile (\ref{flat}) corresponds to the special case $S(z)=z$ with $\Lambda=0$.

Schwarz functions are most commonly defined for closed analytic curves  and much is known about their properties and applications \citep{Davis}.
As shown by \cite{Recent},
it turns out  that
the generalized notion of the Schwarz function $S(z)$ of a wave profile can be used to
express the two-dimensional velocity field
$(u,v)$ associated with steadily-travelling waves with constant vorticity $\omega_0$, and allowing also for submerged cotravelling
periodic rows of point vortices,
in the complex variable form
\begin{equation}
u - {\rm i} v 
=
 - {{\rm i} \omega_0 \over 2} \overline{z}+ q \sqrt{S'(z)} +{{\rm i} \omega_0 \over 2} S(z),
 \label{uvnew}
\end{equation}
where 
$(u,v)$ refers to the velocity field in the cotravelling frame of reference with speed $U_f$, say.
The constant $q$ represents the speed of the fluid on the interface itself; this surface speed must be constant
if both gravity and surface tension are ignored and if the region above the fluid region is at constant pressure.
Condition (\ref{ffc}) means that, as $y \to -\infty$,
\begin{equation}
u - {\rm i} v 
=
 {{\rm i} \omega_0 \over 2} (z- \overline{z} + {\rm i} \Lambda) + q = - {\omega_0 y} + \left (q- {\Lambda \omega_0 \over 2} \right )
\end{equation}
so the wave speed is related to the other parameters via
\begin{equation}
U_f= {\Lambda \omega_0 \over 2} - q.
\label{Uval}
\end{equation}
In view of the general expression (\ref{uvnew}), three cases naturally arise:
\begin{itemize}
\item{Case 1:} $\omega_0 \ne 0, q=0$;

\item{Case 2:} $\omega_0 = 0, q \ne 0$;

\item{Case 3:} $\omega_0 \ne 0, q \ne 0$.

\end{itemize}
In each case,
only special classes of wave profiles will correspond to physically admissible steadily-travelling equilibria, and this means
only special choices of $S(z)$ are allowed in the expression (\ref{uvnew}). 
\cite{Recent}
 shows how to use conformal mapping theory to find
 admissible Schwarz functions and, consequently, to construct
new analytical solutions to the problem of steadily-travelling water waves with vorticity
when the distribution is uniform possibly with superposed  point vortices.
Other singularity types can easily be admitted too.

Using similar conformal mapping techniques
the aforementioned study of
\cite{Nelson} found exact solutions for travelling waves on a deep-water  linear shear current having constant vorticity 
and with a single submerged cotravelling point vortex row.
Their techniques
were borrowed from an earlier study of \cite{crowdy:exactmultipolar}
who posed that streamfunctions taking the form of so-called modified Schwarz potentials
can provide equilibrium vortical solutions of the incompressible two-dimensional Euler equations.
These have the form
\begin{equation}
\psi(z, \overline{z}) = - {\omega_0 \over 4} \left [ z \overline{z} - \int^z S(z') dz' - \overline{\int^z S(z') dz' }
\right ]
\label{Str1}
\end{equation}
for which a simple calculation, with $u =\partial \psi/\partial y, v= -\partial \psi/\partial x$, leads to
\begin{equation}
u - {\rm i} v= 2 {\rm i}{\partial \psi \over \partial z} = - {{\rm i} \omega_0 \over 2} \left ( \overline{z} - S(z) \right )
\label{case1}
\end{equation}
which coincides with (\ref{uvnew}) once the case 1 choice of $q=0$ is made.
It is in this way that
the solutions of \cite{Nelson} can now be viewed as
the most basic water-wave solutions falling within the case 1 category of solutions.

In fact, the framework of \cite{Recent} provides a theoretical unification of 
three (until now, apparently unrelated) contributions in the water-wave literature:
that of
\cite{Nelson}, \cite{Roenby},
and \cite{Miles} 
which, respectively, are now understood as
 the most basic water-wave solutions falling within cases 1, 2 and 3. 
 Interested readers are referred to \cite{Recent} for a more detailed explanation of these developments.
 
 For present purposes it is enough to point out that,
after describing the general framework, \cite{Recent} focussed on producing a range of new solutions falling
within the case 2 category.
Among these are solutions describing two submerged  vortex
rows, also known as von K\'arm\'an vortex streets, cotravelling with a free surface wave
but where the flow was otherwise irrotational;
the earlier work of \cite{Roenby} 
had  found steady waves
cotravelling with a single submerged point vortex row.
 The purpose of the present paper is to 
 present the ``case 1 analogues'' of those new solutions involving two vortex rows: here we present analytical solutions for
submerged
von K\'arm\'an vortex streets (i.e. two vortex rows) cotravelling in a linear shear current beneath a free surface wave thereby generalizing
the work of \cite{Nelson} who focussed on a single cotravelling vortex row.

The paper is set out as follows. In \S \ref{case1background} the background on steady equilibria falling within 
case 1
of the solution taxonomy of \cite{Recent} is given. \S \ref{classic} then describes the classical 
von K\'arm\'an vortex streets in unbounded irrotational flow and examines whether those equilibria
can be generalized to exist in a background simple shear.
In \S \ref{shearcurrent} the problem of two submerged vortex rows, or a vortex street, in a linear
shear current is formulated. It is shown that finding equilibria within the case 1 category can be reduced to the
study of
two algebraic equations whose solution structure is discussed in detail in \S \ref{sec:solve}.
A  characterization of the physically admissible solutions is surveyed in \S \ref{sec:results}.
The paper closes with a discussion of the results in \S \ref{sec:discussion}. 

\section{Case 1 category of solutions \label{case1background}}

Once the expression (\ref{uvnew})
for the complex velocity field has been derived in terms of the Schwarz function $S(z)$ of the wave profile,
the case 1 category of solutions follows simply by taking $q=0$ which means that
the form of the complex velocity field reduces to (\ref{case1}), as explained above. 
This is the generalized viewpoint espoused by \cite{Recent}.
However, because the present paper focusses only on case 1 solutions,
it is
possible to
defer to the earlier work of \cite{Nelson} and offer a more direct formulation in this case.

A vortex patch is the name given to a region of uniform vorticity \citep{PGS}; an unbounded fluid region
of constant vorticity below some wave profile can therefore be viewed as a vortex patch of infinite extent.
For any steadily-travelling wave on the boundary of a vortex patch there is a kinematic
condition that the vortex jump at the patch boundary in a cotravelling
frame of reference must be a streamline.
An additional dynamical condition at the vortex jump says that the velocity fields must be continuous
there: this turns out to ensure the continuity of the fluid pressure \citep{PGS}.

Suppose now that a streamfunction for a steadily-travelling equilibrium over a semi-infinite linear shear layer is given, in a cotravelling
frame, by (\ref{Str1}). It is readily checked that the free surface is a streamline since, on the vortex jump where $\overline{z}=S(z)$,
\begin{equation}
d \psi = {\partial \psi \over \partial z} dz +{\partial \psi \over \partial \overline{z}} d\overline{z} = 
- {\omega_0 \over 4} \left [ \left (\overline{z} - S(z) \right ) dz+ \left ({z} - \overline{S(z)} \right )  d \overline{z} \right ]
=0.
\end{equation}
Moreover since, from (\ref{case1}), $u-{\rm i}v =0$ on the vortex jump then it is continuous with the vanishing
velocity in the upper constant pressure phase.
The streamfunction (\ref{Str1}) therefore appears to furnish a relative equilibrium of the two-dimensional
Euler equations even before any choice of $S(z)$ is made because both the kinematic and dynamic boundary conditions at the vortex jump are satisfied.

The catch is that generic Schwarz functions $S(z)$  have singularities in the region corresponding to the fluid and, as such,
only certain choices of $S(z)$ will be physically admissible. Even then, if $S(z)$ has a physically admissible singularity -- such as a simple
pole with a real residue which corresponds to a point vortex -- there are additional dynamical constraints that any such point vortex is
also in equilibrium with respect to the global configuration.
While all these constraints might appear, at first sight, to 
render it unlikely that equilibrium streamfunctions within this class exist, many such solutions
have now been found.
In the radial geometry most relevant when studying finite-area vortices, such equilibrium
solutions have been identified in
\cite{crowdy:exactmultipolar}, \cite{CroArray}, \cite{crowdy:construction}, \cite{Grow} and \cite{crowdymar}.
For the water wave geometry, the aforementioned work of \cite{Nelson} provides such solutions: that study focussed on
a single submerged point vortex row cotravelling with a wave on the vortex jump on a semi-finite shear layer.
The aim of the present paper is to extend the latter class of water wave solutions to the case
where a submerged von K\'arm\'an vortex street -- that is, a pair of vortex rows, either symmetric (``inline'') or asymmetric (``staggered'') 
\citep{PGS, acheson} --
cotravels with a wave on the vortex jump.

Since the aim here is to study solutions in which vortex streets resembling those studied by
von K\'arm\'an are cotravelling with a wave in a linear shear current,
it is appropriate to review the theory of
von K\'arm\'an vortex streets, without any background shear, in an unbounded 
irrotational flow.

\section{The classical von K\'arm\'an vortex streets \label{classic}}

The complex potential $w(z)$ say, for a single periodic
point vortex row comprising vortices all having circulation $\Gamma$ vortices and
with period $c$ is
well-known \citep{PGS, acheson} to be
\begin{equation}
w(z) = - {{\rm i} \Gamma \over 2\pi} \log \sin \left ({\pi z \over c} \right ),
\end{equation}
where one of the vortices has been placed at the origin.
Apart from the point vortices, the flow is otherwise irrotational.
The associated complex velocity field is
\begin{equation}
u - {\rm i}v = {dw \over dz} = 
- {{\rm i} \Gamma \over 2c} \cot \left ({\pi z \over c} \right ) \to \mp {\Gamma \over 2c}, \qquad
{\rm as}~y \to \pm \infty.
\end{equation}
Far from the vortex row, the fluid velocity
is uniform in the $x$ direction but in opposite directions above and below.

A staggered, or asymmetric, von K\'arm\'an vortex street 
is made up of two such point vortex rows, one with vortices of circulation $\Gamma_u$
above another row with vortices of circulation $\Gamma_l$  offset by half a period; in the classical
setting, $\Gamma_l = - \Gamma_u$ for reasons to be seen shortly.
Such a street, with period $c=2 \pi$,  therefore
has complex potential
\begin{equation}
- {{\rm i} \Gamma_u \over 2\pi} \log \sin \left ({z+{\rm i} \over 2} \right )
- {{\rm i} \Gamma_l \over 2\pi} \log \sin \left ({z - \pi + {\rm i} (1+\lambda)  \over 2} \right ),
\end{equation}
where
we have now placed one of the vortices in the row 
having circulation $\Gamma_u$ at $z=-{\rm i}$
and 
 one of the vortices in the row having circulation $\Gamma_l$ at $z=\pi-{\rm i}(1+\lambda)$;
 this is for ease of comparison with solutions found later.
 The parameter $\lambda$ is the aspect ratio of the street \citep{PGS, acheson}.
 The associated complex velocity field is 
 \begin{equation}
u-{\rm i} v = - {{\rm i} \Gamma_u \over 4\pi}  \cot \left ({z+{\rm i} \over 2} \right )
- {{\rm i} \Gamma_l \over 4\pi}  \cot \left ({z -\pi+ {\rm i} (1+\lambda)  \over 2} \right ).
\end{equation}
Since both cotangent functions tend to $\mp {\rm i}$ as $y \to \pm \infty$,
it is clear that the velocity induced far away from this street will only vanish provided
$\Gamma_u = -\Gamma_l:=\Gamma$. It is then easy to show, using the usual rules
for the velocity of a free point vortex, that the vortex street moves steadily
in the $x$ direction with
velocity
\begin{equation}
 {\Gamma \over 4\pi}  \tanh \left ({ \lambda  \over 2} \right ).
\end{equation}

It is natural to ask whether such a relative equilibrium can also exist
if placed in a simple shear flow, $(-y, 0)$ say.
There is no complex potential in this case, but
the associated complex velocity field is
 \begin{equation}
u-{\rm i} v = -y - {{\rm i} \Gamma \over 4\pi}  \cot \left ({z+{\rm i} \over 2} \right )
+{{\rm i} \Gamma \over 4\pi}  \cot \left ({z -\pi + {\rm i} (1+\lambda)  \over 2} \right ),
\label{VK}
\end{equation}
where the same relationship $\Gamma_u = -\Gamma_l:=\Gamma$ is again necessary
to ensure there is no uniform flow component in the far-field.
Suppose we assume the existence of a steadily-translating equilibrium
moving in the $x$ direction with speed $U$. Then the condition
 for equilibrium 
at $y=-1$, or $z=-{\rm i}$, is
 \begin{equation}
U_{\rm stag} = 1 +
 {\Gamma \over 4\pi}  \tanh \left ({ \lambda  \over 2} \right )
 \label{condition1}
\end{equation}
while
 the condition
at $y=-(1+\lambda)$, or $z=-{\rm i}(1+\lambda)$, is
 \begin{equation}
U_{\rm stag} = 1+\lambda  +  {\Gamma \over 4\pi}  \tanh \left ({ \lambda  \over 2} \right ).
 \label{condition2}
\end{equation}
It is clear that (\ref{condition1}) and (\ref{condition2}) can only be consistent if $\lambda=0$, corresponding
to a degenerate case, with zero aspect ratio, comprising
a periodic row of vortices of alternating circulation spaced apart by $\pi$.

A similar conclusion is reached for the inline (also known as unstaggered, or symmetric) 
von K\'arm\'an vortex streets. In this case, provided $\lambda \ne 0$ (because otherwise
the two point vortex rows will sit directly atop each other and cancel each other out),
the analogues of the two conditions
(\ref{condition1}) and (\ref{condition2}) are
 \begin{equation}
U_{\rm inline} = 1 +
 {\Gamma \over 4\pi}  \coth \left ({ \lambda  \over 2} \right )
 \label{condition3}
\end{equation}
and
 \begin{equation}
U_{\rm inline} = 1+\lambda  +  {\Gamma \over 4\pi}  \coth \left ({ \lambda  \over 2} \right ).
 \label{condition4}
\end{equation}
Since $\lambda=0$ is the only consistent solution of both (\ref{condition3}) and (\ref{condition4}), and
 because this value corresponds to the two 
 vortices cancelling each other out,
the conclusion is that 
there is no equilibrium for an inline, or symmetric, 
von K\'arm\'an vortex street in an unbounded simple shear current.

These simple calculations reveal that only degenerate cases of the classical von K\'arm\'an vortex street
equilibria survive when placed in an unbounded simple shear.
Interestingly, however, in what follows we are able to show that 
equilibria resembling
von K\'arm\'an vortex streets
 do survive when  the point vortices are cotravelling with a free surface wave
in a simple shear current.


\section{von K\'arm\'an vortex streets cotravelling with a wave in a linear shear current \label{shearcurrent}}

\begin{figure}
  \centering
  \includegraphics[scale=0.5]{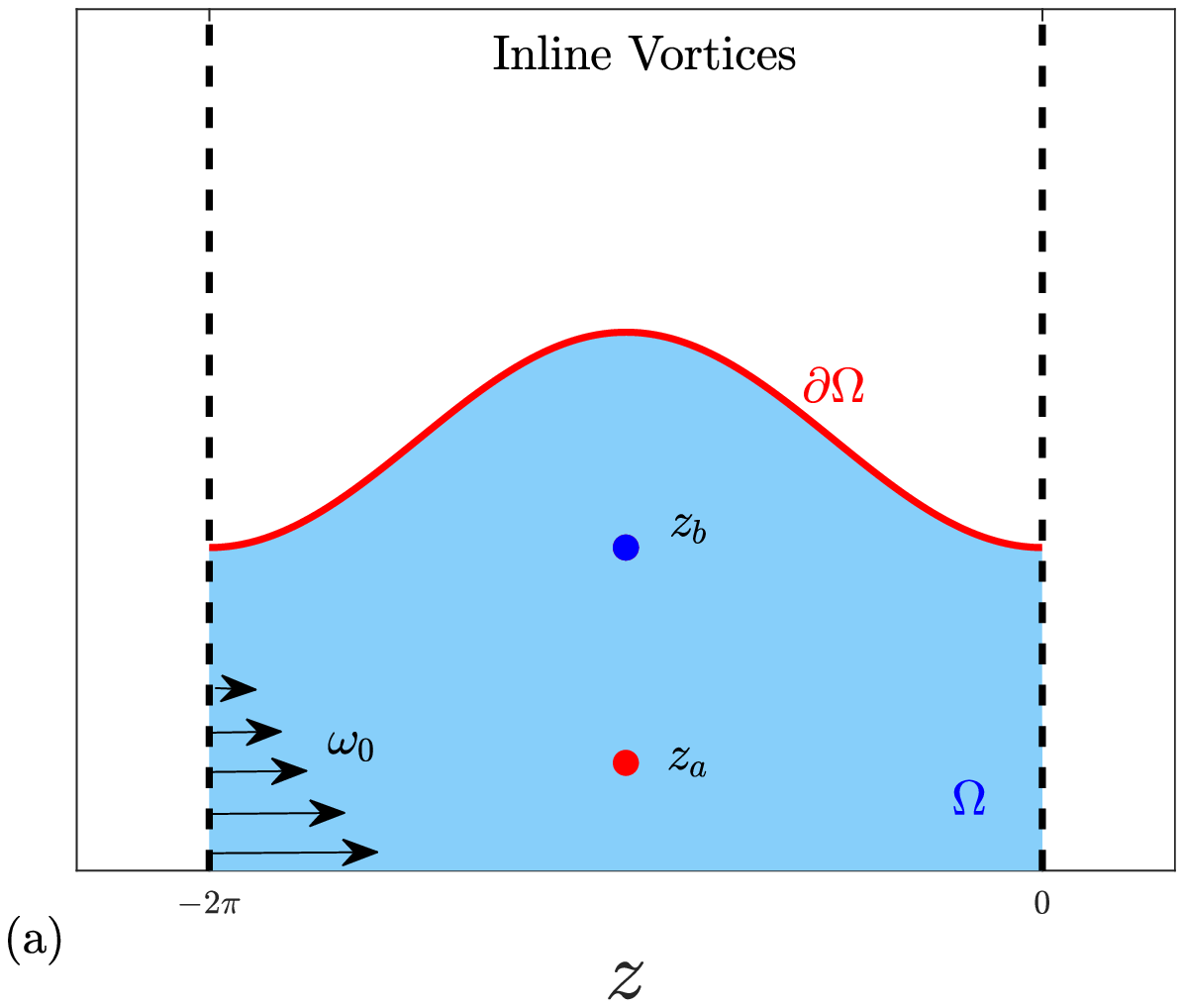}
  \includegraphics[scale=0.5]{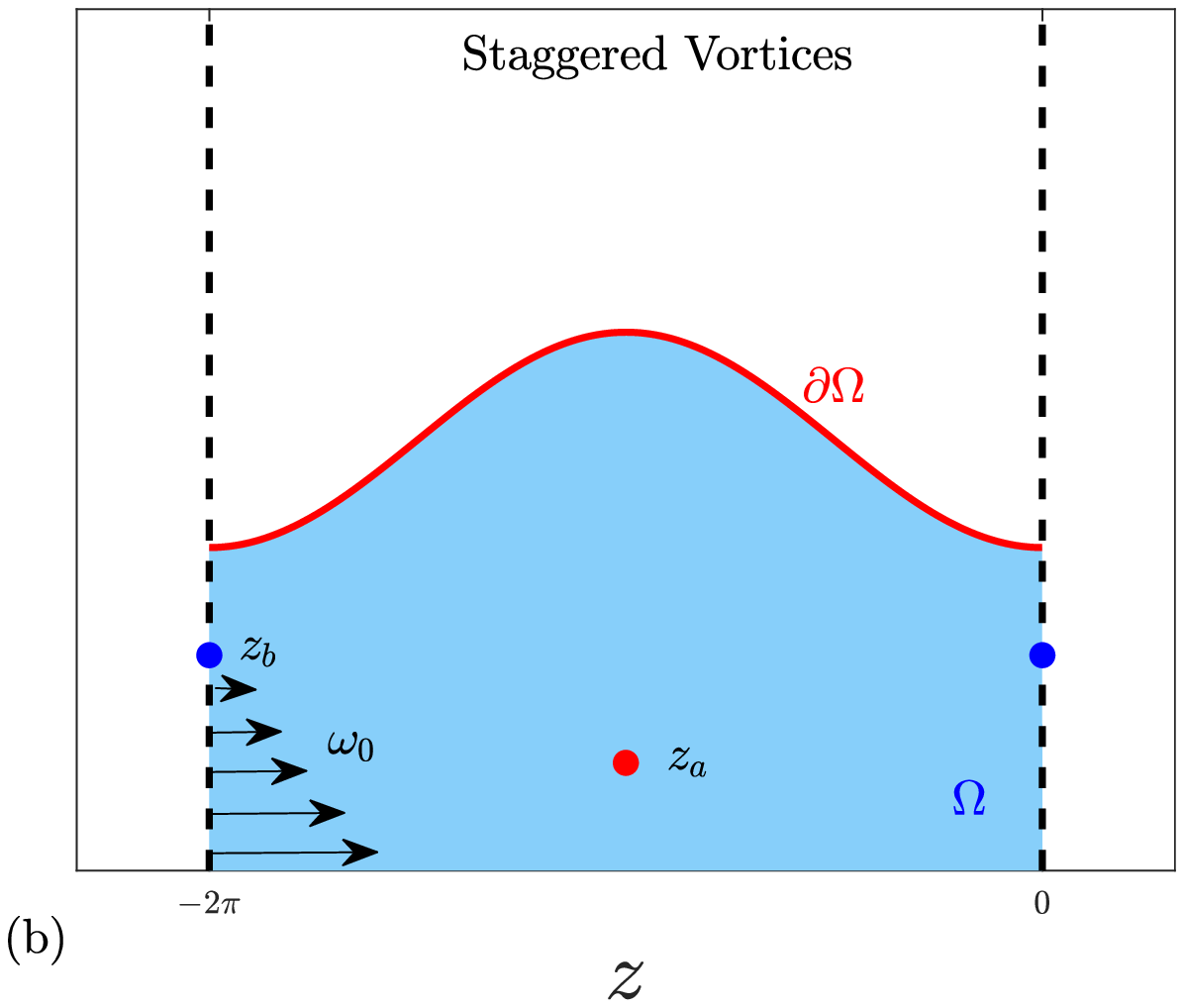}
  \caption{The physical plane for (a) inline and (b) staggered vortices. The position of the vortices are denoted $z_a$ and $z_b$ and the domain is 2$\pi$-periodic.}
  \label{fig:physical}
\end{figure}

The physical situations
of interest for the remainder of this paper are
  shown in figure~\ref{fig:physical} which shows (a) inline configurations and (b) staggered configurations in a single periodic window of the complex $z$ plane. 
  The fluid domain in this period window 
  is denoted by $\Omega$: it is unbounded as $y \to -\infty$ but bounded above by a free 
surface, denoted by $\partial\Omega$.
 As $y \to -\infty$ the flow tends to a linear shear of the form $(-y,0)$ which
 requires the choice $\omega_0=1$; this simply sets a timescale for the flow.
 It is assumed that there are two point vortices 
  in the fluid in each period window. 
  For inline configurations both are located in the middle of the period window; 
  for the staggered configuration the vortices are offset by half a period in the $x$-direction.
  Since steadily-travelling equilibria with speed $U_f$ in the positive $x$-direction are sought, it is natural
  to move to a cotravelling frame of reference where, as
    $y \to -\infty$, the flow is steady and tends to a linear shear of the form $(-y-U_f,0)$. 
    In this frame of reference, the wave profile is fixed, so too are the locations of any submerged point vortices.
    This means, according to the usual equations of motion of a free point vortex \citep{PGS, acheson}, 
    that the non-singular component of the velocity field at each point vortex must vanish.
  
Following the formulation in \cite{Nelson} who allowed for a single point vortex
per period, or a single submerged vortex row, the extension to two point vortices per period,
or two submerged vortex rows,
requires consideration of a generalized conformal map of the form
\bea
z = Z(\zeta) = \mathrm{i}\left[\log \zeta + \frac{A}{\zeta - a} + \frac{B}{\zeta - b}\right] + d,
\label{conformal_map}
\eea
where $a, b, A, B \in \mathbb{R} $ and $d \in {\rm i} \mathbb{R}$ are parameters
to be determined.
This mapping
transplants a unit disc, in a parametric $\zeta$ plane, to the period window 
$\Omega$ in the $z$-plane. 
Let the interior of the unit disc be denoted by $D$ and its unit-circle
boundary by $\partial D$. 
It is necessary that $|a|,|b|>1$ to ensure the there are no poles of $z$ in $D$; this is
because the conformal mapping must be analytic in the disc except
for the logarithmic singularity at $\zeta=0$ which is required by the periodic nature
of the image domain.
 The boundary $\partial D$ is transplanted to the free
 surface $\partial \Omega$ in the physical plane. The two sides of a logarithmic branch
 cut between $\zeta=0,\infty$ are transplanted to the two sides of the period window $\Omega$.

\begin{figure}
  \centering
  \includegraphics[scale=0.5]{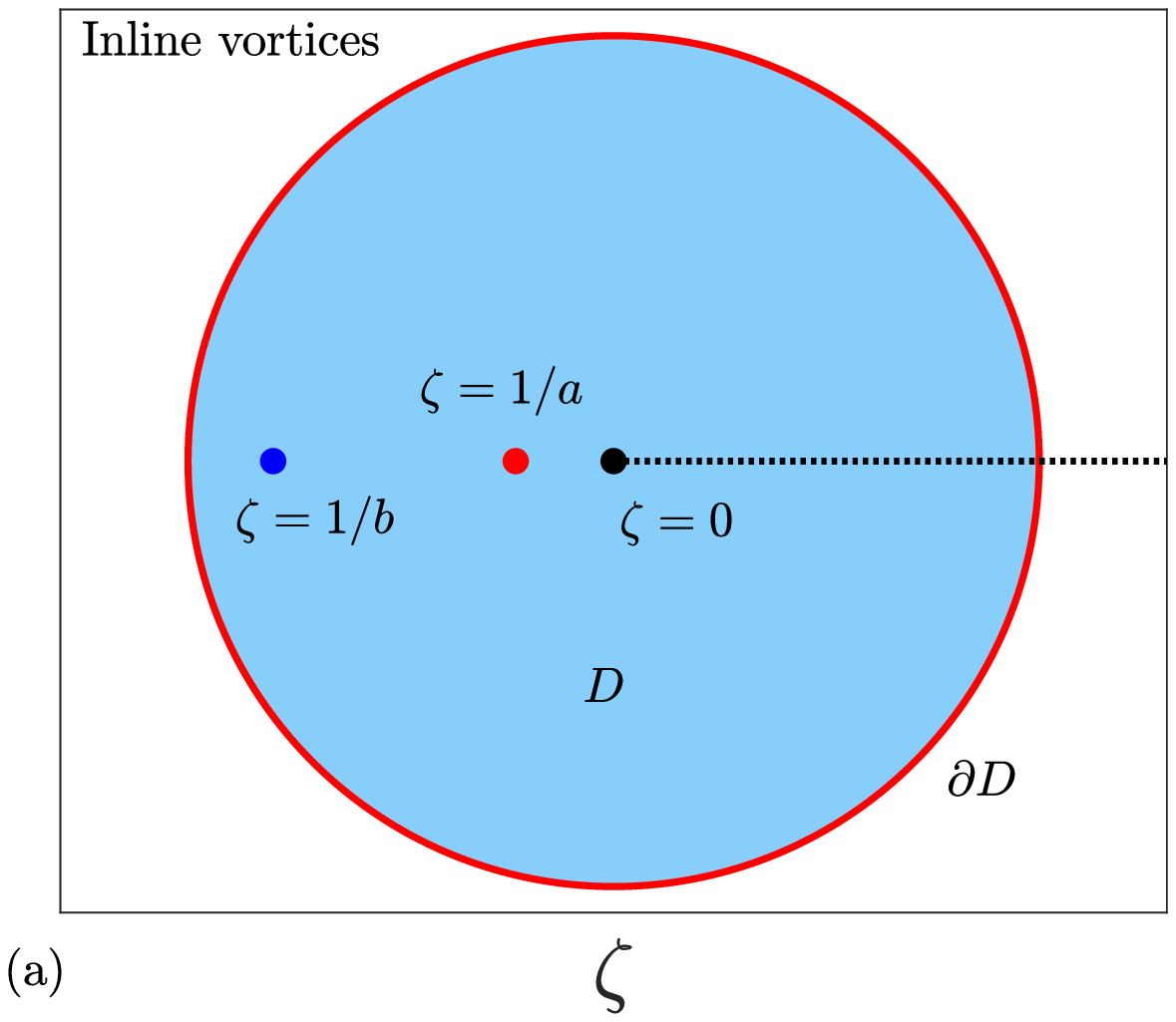}
  \includegraphics[scale=0.5]{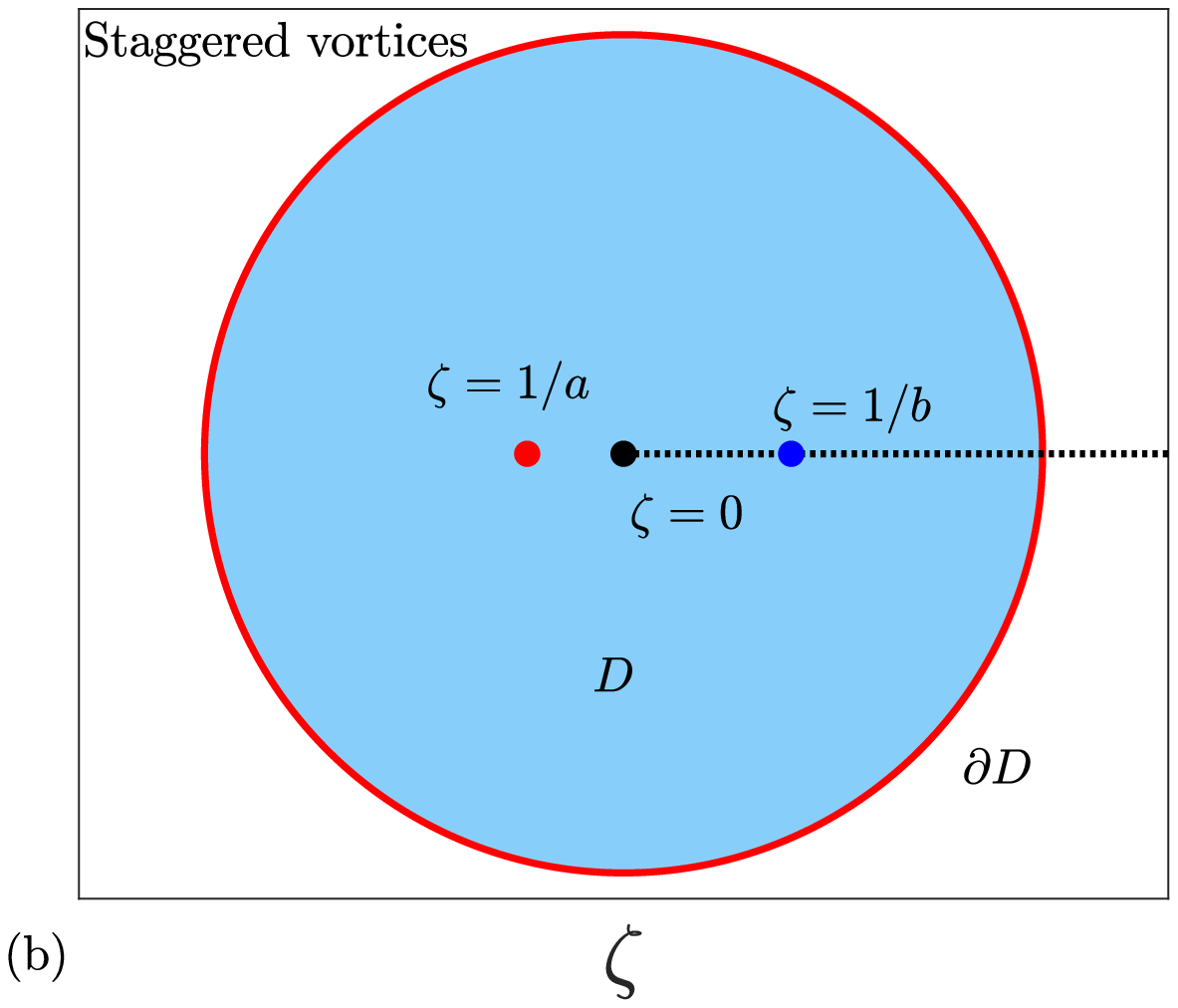}
  \caption{The $\zeta$ plane for (a) inline and (b) staggered vortices. The branch cut is shown as a dashed line. The singularities of $\textbf{u}$ are at $\zeta = 1/a$ and $\zeta = 1/b$ and correspond to the positions of the vortices in the $\zeta$ plane.}
  \label{fig:disc}
\end{figure}

To see how \eqref{conformal_map} produces the two point vortices per period
note that
the Schwarz function can be written, as a function of $\zeta$, as follows:
\bea
S(z) = 
\overline{z} =  \overline{Z(\zeta)} = \overline{Z}(1/\zeta) = 
\mathrm{i}\left[\log \zeta - \frac{A\zeta}{1 - a\zeta} - \frac{B\zeta}{1 - b\zeta}\right] + \overline{d},
\eea
where we have used the fact that $\overline{\zeta} =1/\zeta$ on $\partial D$, and hence on $\partial \Omega$.
Since this function has
the same logarithmic singularity as $Z(\zeta)$ at $\zeta=0$
it is easy to check that this function satisfies the far-field condition (\ref{ffc}).
The Schwarz function
$S(z)$ has simple
poles at $\zeta = 1/a,1/b$, which are inside $D$ and therefore correspond to
simple poles of $S(z)$ at
\begin{equation}
\begin{split}
z_a &= Z(1/a) = \mathrm{i}\left[\log(a^{-1}) + \frac{Aa}{1 - a^2} + \frac{Ba}{1 - ab}\right] + d
\\ {\rm and}~~z_b &= Z(1/b) = \mathrm{i}\left[\log(b^{-1}) + \frac{Bb}{1 - b^2} + \frac{Ab}{1 - ab}\right] + d
\end{split}
\end{equation}
where are inside $\Omega$. 
%
The two parameters $a$ and $b$ will be viewed as free parameters.
It then turns out that, for equilibrium, $A = A(a,b)$, $B=B(a,b)$ and $d=d(a,b)$ must be determined
as functions of these two parameters.
To see this, notice that
the complex velocity field (\ref{case1})
can be written, as a function of $\zeta$, as
\begin{equation}
u - {\rm i}v = -{{\rm i} \omega_0 \over 2} \left [ \overline{z}- S(z) \right ]
=
-{\omega_0 \over 2} \left [
\log |\zeta|^2 + {A \over \overline{\zeta}-a} + {B \over \overline{\zeta}-b}
-{A \zeta \over 1- \zeta a} - {B \zeta \over 1 -\zeta b} \right ].
\end{equation}
To find the condition that the point vortex at $\zeta=1/a$ is in equilibrium, it is useful to rewrite
the velocity field as
\begin{equation}
u - {\rm i}v 
=
-{\omega_0 \over 2} \left [
\log |\zeta|^2 + {A \over \overline{\zeta}-a} + {B \over \overline{\zeta}-b}
+{A \over a} + {A \over a^2} {1 \over (\zeta-1/a)} +{B \zeta \over 1 -\zeta b} \right ]
\end{equation}
and then make use of the fact that, near $\zeta=1/a$,
\begin{equation}
{1 \over \zeta-1/a} = {Z'(1/a) \over z-z_a} + {Z''(1/a) \over 2 Z'(1/a)} + {\cal O}(z-z_a).
\label{Routh}
\end{equation}
It follows that, near $z=z_a$,
\begin{equation}
u - {\rm i}v 
=
-{\omega_0 \over 2} \left [ {A \over a^2} {Z'(1/a) \over z-z_a} +
 S_a + {\cal O}(z-z_a) \right ],
\end{equation}
where
\begin{equation}
S_a = \log(1/a^2)+ {A a \over 1-a^2} +{Ba \over 1-ab} +{A \over a} +
{A \over a^2} {Z''(1/a) \over 2 Z'(1/a)} + {B \over b-a}.
\end{equation}
Thus a point vortex, of circulation $\Gamma_a$ at $z=z_a = Z(1/a)$, where
\begin{equation}
-{{\rm i} \Gamma_a \over 2\pi} = -{A \omega_0 Z'(1/a) \over 2a^2}
\label{circA}
\end{equation}
will be in equilibrium provided that $S_a=0$.
This is the usual equilibrium condition for a free point vortex.
By exactly the same reasoning, the
point vortex, of circulation $\Gamma_b$ at $z=z_b= Z(1/b)$ where
\begin{equation}
-{{\rm i} \Gamma_b \over 2\pi} = -{B \omega_0 Z'(1/b) \over 2b^2}
\label{circB}
\end{equation}
will be in equilibrium provided that $S_b=0$ where
\begin{equation}
S_b =\log(1/b^2)+ {A b \over 1-b^2} +{Bb\over 1-b^2} +{B \over b} +
{B \over b^2} {Z''(1/b) \over 2 Z'(1/b)} + {A \over a-b}.
\end{equation}
The two equilibrium conditions can be rewritten as
\begin{equation}
\begin{split}
2a^2 \left [ \log(1/a^2) +{A \over a(1-a^2)} + {B(1-a^2) \over (1-ab)(b-a)} \right ] Z'(1/a)
+ AZ''(1/a) &= 0, \\
2b^2 \left [ \log(1/b^2) + {A(1-b^2) \over (1-ab)(a-b)} +{B \over b(1-b^2)} \right ] Z'(1/b) + B Z''(1/b) &=0.
\end{split}
\end{equation}
The algebraic form of these equations is 
\begin{equation}
  \begin{split}
  \lambda_1A^2 + \lambda_2B^2 + \lambda_3AB + \lambda_4A + \lambda_5B + \lambda_6 &= 0  \\
  \mu_1A^2 + \mu_2B^2 + \mu_3AB + \mu_4A + \mu_5B + \mu_6 &= 0,  
  \end{split}
  \label{conicA}
  \end{equation}
where the coefficients $\lbrace \lambda_j, \mu_j | j=1, \dots, 6 \rbrace$ are given
as explicit functions of $a$ and $b$ in 
appendix \ref{appA}.
These equations will be viewed as determining $A$ and $B$ as functions
of $a$ and $b$, i.e., $A = A(a,b)$, $B=B(a,b)$. Since
it determines the possible equilibria, 
the solution structure of (\eqref{conicA}) is discussed in detail in the following section,
not least because it is found to have intriguing and unexpected features.

It only remains to fix $d$. But
 we are free to set the location of the vortex at $z_a$ so, following \cite{crowdy2010steady}, 
 we set $z_a = -\pi - \mathrm{i}$, i.e. at unit distance below $y=0$
 in the middle of the period window. This determines $d$.
 An alternative choice is to pick $d$ so that mean level of the wave profile is specified,
 but then we would lose control of the position of one of the vortices.

\section{The solution structure of (\ref{conicA}) for $A$ and $B$}\label{sec:solve}


The possibility of
finding equilibria has been reduced to
 determining $A$ and $B$ from the algebraic system \eqref{conicA}. For a single vortex, the equivalent single equation can easily be manipulated into a closed-form expression that gives $A$ as a straightforward function of $a$ \citep{crowdy2010steady}. In the present two-vortex case 
 the analysis is more involved but, as will be discussed here, it shares some
 interesting and surprising features.
 Due to the complicated nature of the coefficients $\lambda_i,\mu_i$, although an analytic solution is possible in principle, it is too cumbersome to gain insight.

%


\begin{figure}
  \centering
\includegraphics[width=\textwidth]{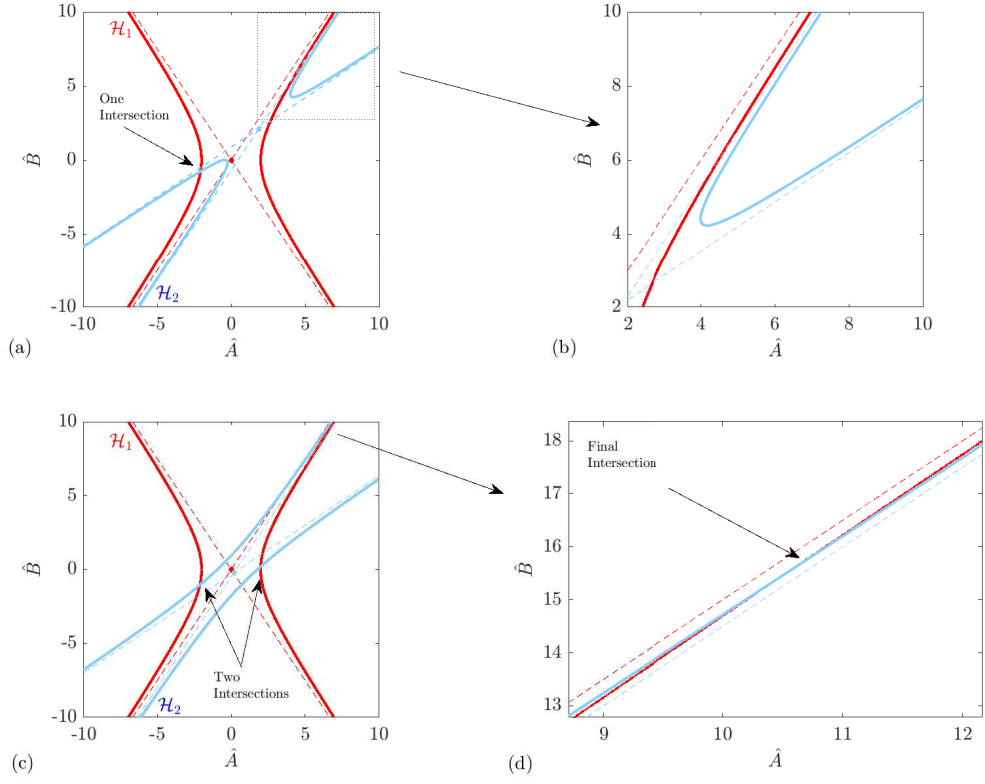}
\caption{(a),(b) Two hyperbolae with the properties described in \eqref{properties} that intersect at 1 point. (b) A zoom-in of the rectangular area in (a) showing no further intersection is possible. (c),(d) Two hyperbolae with the properties described in \eqref{properties} that intersect at 3 points. (d) A zoom-in near to where the third intersection occurs.}
\label{fig:hyperbolae}
\end{figure}

\begin{figure}
  \centering
\includegraphics[scale=0.5,trim = 0 0 0 0,clip]{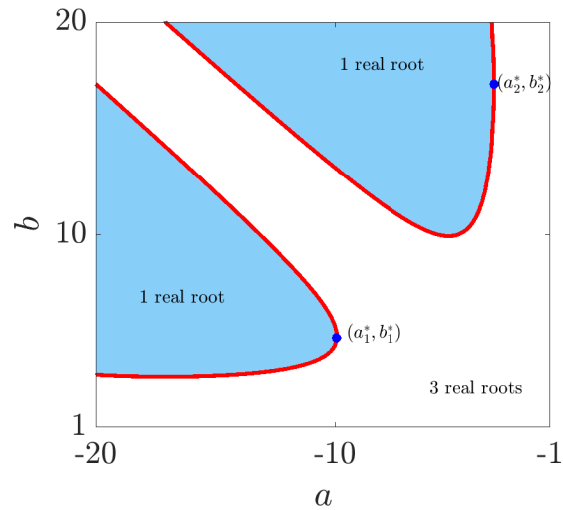}
\caption{Regions in the $(a,b)$ plane which result in 1 real root (shaded) or 3 real roots (non-shaded) for the staggered vortex configuration. The special parameter values $(a_i^*,b_i^*)$, $i=1,2$ (as marked in the figure) correspond to transcritical bifurcations.}
\label{fig:rootregions}
\end{figure}

One would expect
from its general algebraic structure in \eqref{conicA}, and using
 B\'{e}zout's theorem \citep{bezout1779theorie} that, accounting for multiplicity of the roots, there will be four pairs of solutions.\footnote{The number of solutions is the product of the highest power in each equation} Indeed, the solutions, $(A,B)$, represent the intersection of two conic sections.

However, by working through the algebra, 
it turns out that
$\lambda_1 \equiv 0$ and $\mu_2 \equiv 0$;
see appendix \ref{appA}.
 Therefore $\Delta = \lambda_3^2 - 4\lambda_1\lambda_2 > 0$ and thus the two conics are both hyperbolae. One can easily eliminate, say $A$, from the equations in \eqref{conicA} to leave a quartic equation for $B$, which has four roots, as predicted, with closed-form expressions. Remarkably, again working through the algebra, 
it turns out that the coefficient of $B^4$ is zero, namely,
\bea
\mu_1\lambda_2^2 - \mu_3\lambda_2\lambda_3 \equiv 0.
\eea
This degeneracy implies \eqref{conicA} has only three roots; either 1 real and 2 complex, or 3 real.\footnote{We note that this does not contradict B\'{e}zout's theorem as the extra root can be accounted for by the intersection of the two hyperbolae at infinity.} 

It is worth emphasizing that
 a similar phenomenon occurs in the single vortex
 analysis of \citep{crowdy2010steady}:
a single equation for $A$ is, at first glance, a quadratic equation, but the coefficient of the 
leading quadratic term is identically zero, resulting in a one-parameter family of solutions \citep{crowdy2010steady}. More will be said on this observation  in \S~\ref{sec:discussion}.  

It is useful to understand this degenerate case by exploring the geometry of the curves defined in \eqref{conicA}. Examining the large $(A,B)$ behaviour in \eqref{conicA} we find that
\begin{align}
B \sim 0,&\qquad \lambda_2B + \lambda_3A \sim 0\\
A \sim 0,&\qquad \mu_3B + \mu_1A \sim 0.
\end{align}
The asymptotes corresponding to $A\sim0$ and $B\sim0$ are vertical and horizontal lines in the $(A,B)$ plane, respectively, and the gradients of the non-trivial asymptotes are $-\lambda_2/\lambda_3$ and $-\mu_3/\mu_1$, respectively. We find that $-\lambda_2/\lambda_3 = -\mu_3/\mu_1 \equiv (1 - b^2)/(a^2-1)$ and thus the asymptotes are parallel. The problem thus reduces to finding the intersection of two hyperbolae, $\mathcal{H}_1$ and $\mathcal{H}_2$, with the following two properties:
\begin{align}
&\mbox{1)}\qquad \mbox{One asymptote of }\mathcal{H}_1 \mbox{\:is perpendicular to }\mathcal{H}_2\\
  &\mbox{2)}\qquad \mbox{One asymptote of }\mathcal{H}_1\mbox{\:is parallel to }\mathcal{H}_2
  \label{properties}
\end{align}
In figure~\ref{fig:hyperbolae} we sketch a standard rectangular hyperbolae (we can always perform a transformation on one of the conics in \eqref{conicA} to the standard form) with centre $(0,0)$ (red curves) and another hyperbola satisfying the properties in \eqref{properties}. In panels (a) and (b) we see that $\mathcal{H}_1$ and $\mathcal{H}_2$ only intersect at one point, whilst in panels (c) and (d) we show how they can intersect at three points. In panel (a) we also see that the extra `roots' predicted by B\'{e}zout's theorem are accounted for by the two curves `intersecting' at infinity.

This analysis is important because na\"ively solving \eqref{conicA} using a computer algebra package can be expensive, inefficient and can sometimes not even give an answer in the allotted time. In practice, it was found that the most computationally efficient method was to reduce \eqref{conicA} to a single cubic equation for $B$ say, and then apply the cubic formula to find the three roots in terms of $a$ and $b$. Note that there is no way of knowing, \textit{a priori}, that \eqref{conicA} reduces to a cubic equation. A double precision $(a,b)$ meshgrid was constructed and  the roots calculated
using the analytic expressions. 

It should be noted that solutions to \eqref{conicA} are not necessarily \textit{valid} solutions to the physical water-wave problem:
this is because of the additional requirement that the mapping
 \eqref{conformal_map} is a one-to-one, or univalent, mapping
 from $D$ to $\Omega$.
In the case of three pairs of real roots, the solution is non-unique for the given $(a,b)$. However, as we shall discuss in the next section, a valid solution can only be constructed in certain regions of $(a,b)$ parameter space. We can find the regions of $(a,b)$ space where there are 1 or 3 roots by calculating the discriminant of the resulting cubic equation in terms of $a$ and $b$. For inline vortices, i.e. when $ab>0$, we find the discriminant is always positive except when $a=b$ and there is no solution, therefore there are always three real roots. For staggered vortices, $ab<0$, the discriminant can be negative, allowing for a single real solution; figure~\ref{fig:rootregions} indicates these regions in the $(a,b)$ plane. Before discussing these solutions in more detail, it is worth discussing the solution structure in two particular limits, when $a-b\to 0$ and when $a,b\to 1$.

\subsection{The limit $a-b\to 0$}\label{sec:limit1}

In this limit, only applicable to inline configurations, by multiplying both equations in \eqref{conicA} by $a-b$, and then taking the limit $a-b\to 0$, we find that the \eqref{conicA} reduces to
\begin{equation}
  \begin{split}
    \hat{\lambda}_2B^2 + \hat{\lambda}_3AB + \hat{\lambda}_5B &= 0  \\
    \hat{\mu}_1A^2 + \hat{\mu}_3AB + \hat{\mu}_4A &= 0,  
  \end{split}
  \label{conicC}
\end{equation}
where $\hat{\lambda}_i = (a-b)\lambda_i,\hat{\mu}_i = (a-b)\mu_i$. This has a single trivial solution $(A,B) = (0,0)$; the other roots do not exist as $\hat{\lambda}_2\hat{\mu}_3 -  \hat{\lambda}_3\hat{\mu}_1 = 0$. Physically, this limit corresponds to a flat profile with two increasingly close vortices that effectively disappear when $a=b$.

\subsection{The limit $a,b\to -1$}\label{sec:limit2}

In this limit, because (concentrating on $a\to -1$, the limit is $b\to -1$ is similar), if we multiply the terms in \eqref{conicA} by $(a^2-1)^2$ and take the limit as $a\to -1$ we find that all of the coefficients identically vanish. However the dominant behaviour of \eqref{conicA} in this limit is
\begin{equation}
  \begin{split}
    \bar{\lambda}_4A \sim 0, \qquad
    \bar{\mu}_5B \sim 0,  
  \end{split}
  \label{conicD}
\end{equation}
where $\bar{\lambda}_i = (1-a^2)^2\lambda_i,\bar{\mu}_i = (1-b^2)^2\mu_i$. Therefore as $a,b\to -1$, $(A,B)\to 0$.


\section{Characterization of the equilibria }\label{sec:results}

The inline and staggered configurations will be considered separately. 
In each case the solution space for $(A,B)$
is discussed
 as functions of the free parameters $(a,b)$.
In what follows,
a valid solution is defined to be a set of parameters $(a,b)$ with solutions $(A,B)$ of \eqref{conicA} for which the mapping in \eqref{conformal_map} is univalent, i.e. the there are no intersections of the interface in the physical $z$ plane. We shall discuss the conditions of validity as they arise in the analysis.

\subsection{The inline (unstaggered) vortex street}

\begin{figure}
  \centering
  \includegraphics[width=\textwidth]{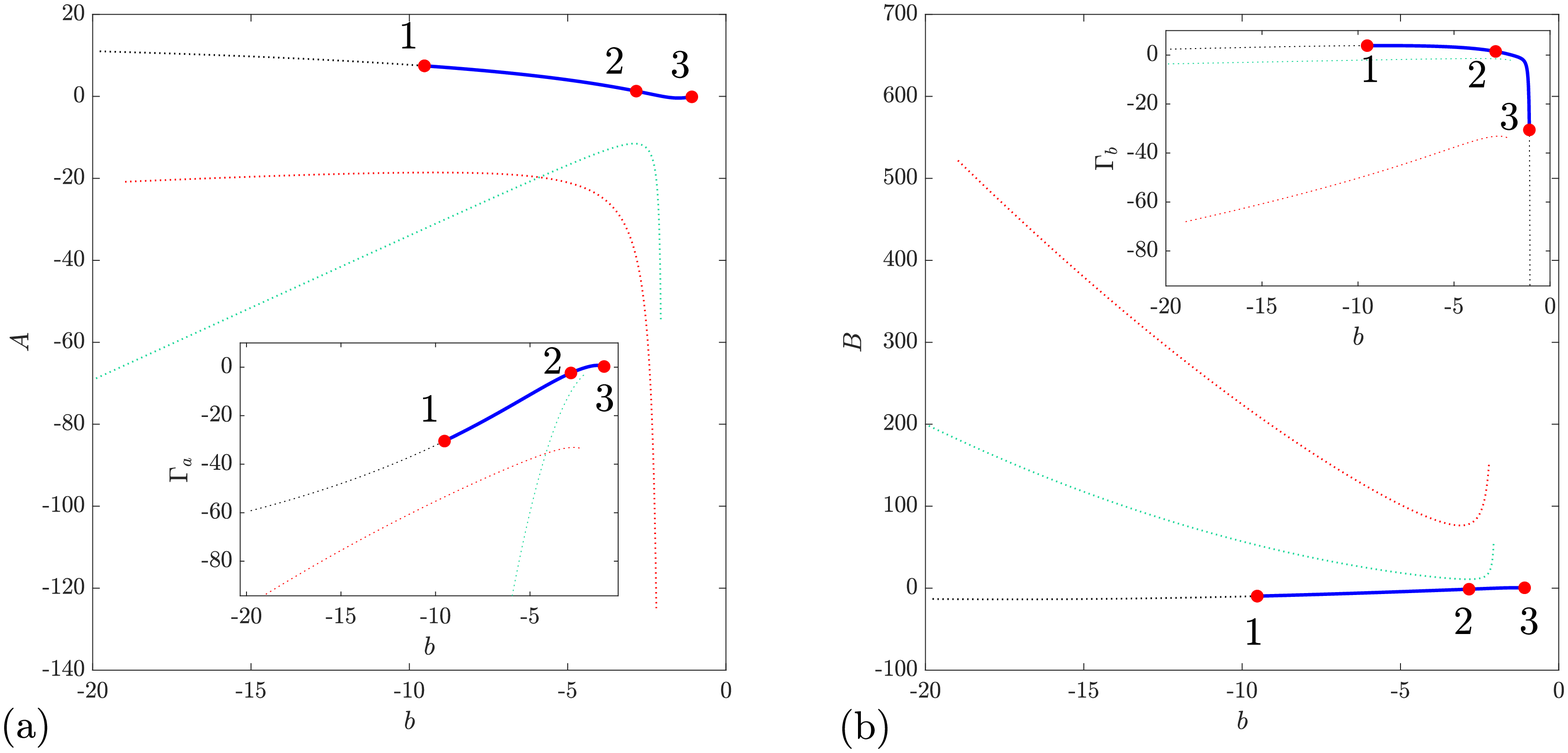}
  \includegraphics[width=\textwidth]{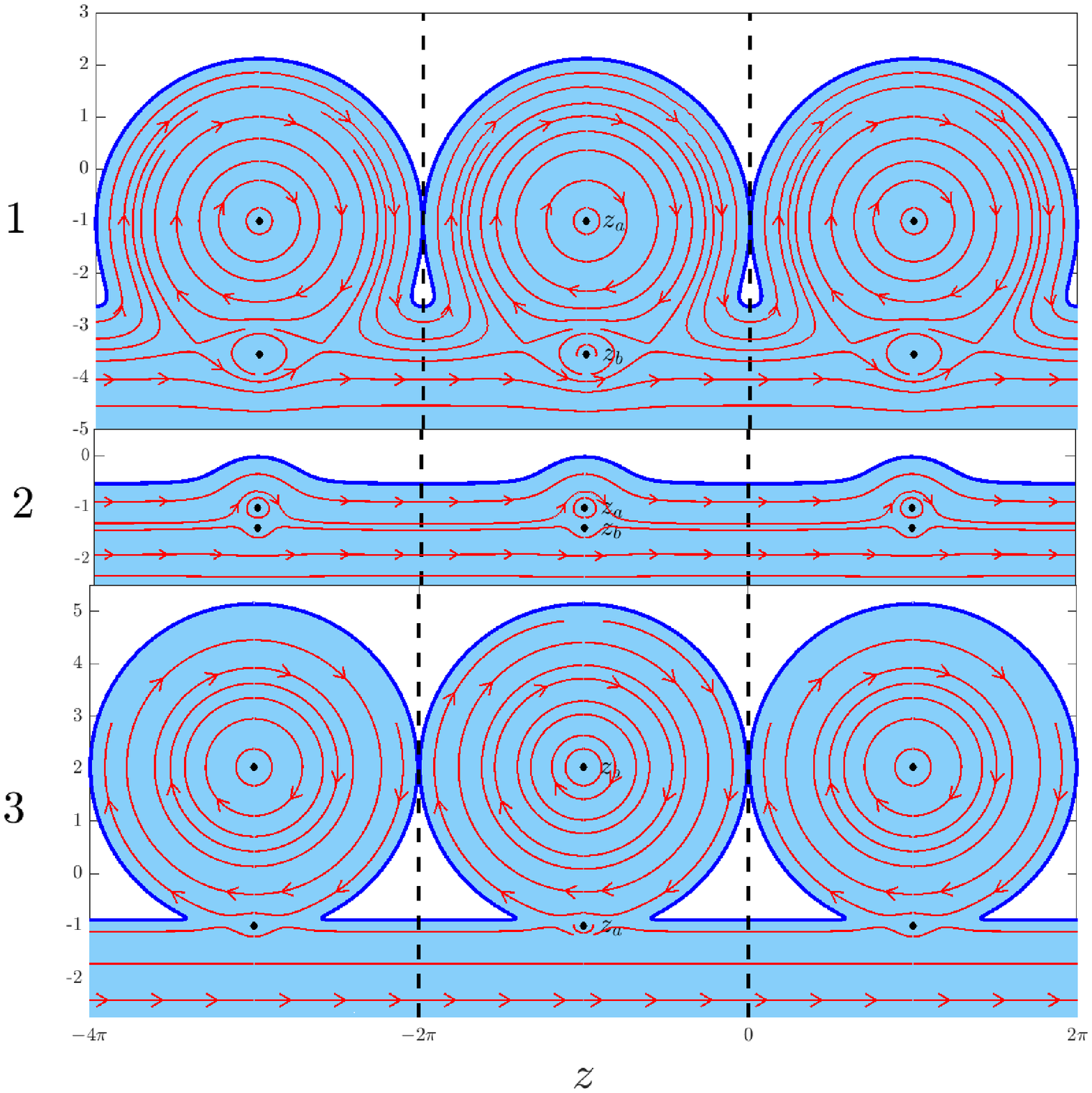}
  \caption{Solution structure for inline vortices with $a=-2$. (a)/(b) shows plots of $A/B$, respectively, as $b$ is varied. The inset diagrams show the variation of the vortex strength, $\Gamma_a$/$\Gamma_b$ in panels (a)/(b) respectively. The solid lines indicate solutions of \eqref{conicA} that result in a univalent mapping. Thin dotted lines indicate solutions of \eqref{conicA} that do not result in a valid solution. The streamlines and profiles are show for the solutions labelled 1,2 and 3 in panel (a) as well as the positions of the vortices.}
  \label{fig:inline_bifurcation}
\end{figure}

As mentioned in the previous section,
inline vortices ($ab>0$) have a positive discriminant of \eqref{conicA} and there are always three pairs of real solutions. Figure~\ref{fig:inline_bifurcation} shows the solutions $A$ and $B$ (panels (a) and (b) respectively) as $b$ is varied when $a=-2$. Each different coloured branch represents one of the roots of \eqref{conicA}, with solid/dotted lines indicating univalent/non-univalent mappings. For $a = -2$, there is only a small portion of one branch that contains univalent mappings and thus represent physical wave profiles. The limiting profiles (profiles 1 and profiles 3) that occur at end of this branch portion both self-intersect with a neighbouring periodic window when $\zeta\neq 1$. The profile labelled 2 indicates a solution which is almost flat in the far-field. We remark that the solution branch crosses the line $a=b$ but no solution exists when $a=b$ exactly. When $b<a$, the $z_a$ vortex is the upper vortex and the vice versa when $b>a$.

The vortex strengths, $\Gamma_a$ and $\Gamma_b$ are plotted in the inset diagrams of panel (a) and (b) respectively. The circulations at the limiting profile, labelled 3, are $\Gamma_a = 0.2711, \Gamma_b = -30.5539$ which indicates the lower $z_a$ vortex has significantly less influence on the flow than the upper $z_b$ vortex. The circulations of the other limiting profile, labelled 1, are $\Gamma_a = -30.4596, \Gamma_b = 3.8487$, so that although the lower $z_b$ vortex has less influence on the flow than the upper $z_b$ vortex, it is not as weak as the lower vortex in profile 3.

\begin{figure}
  \centering
  \includegraphics[width=\textwidth]{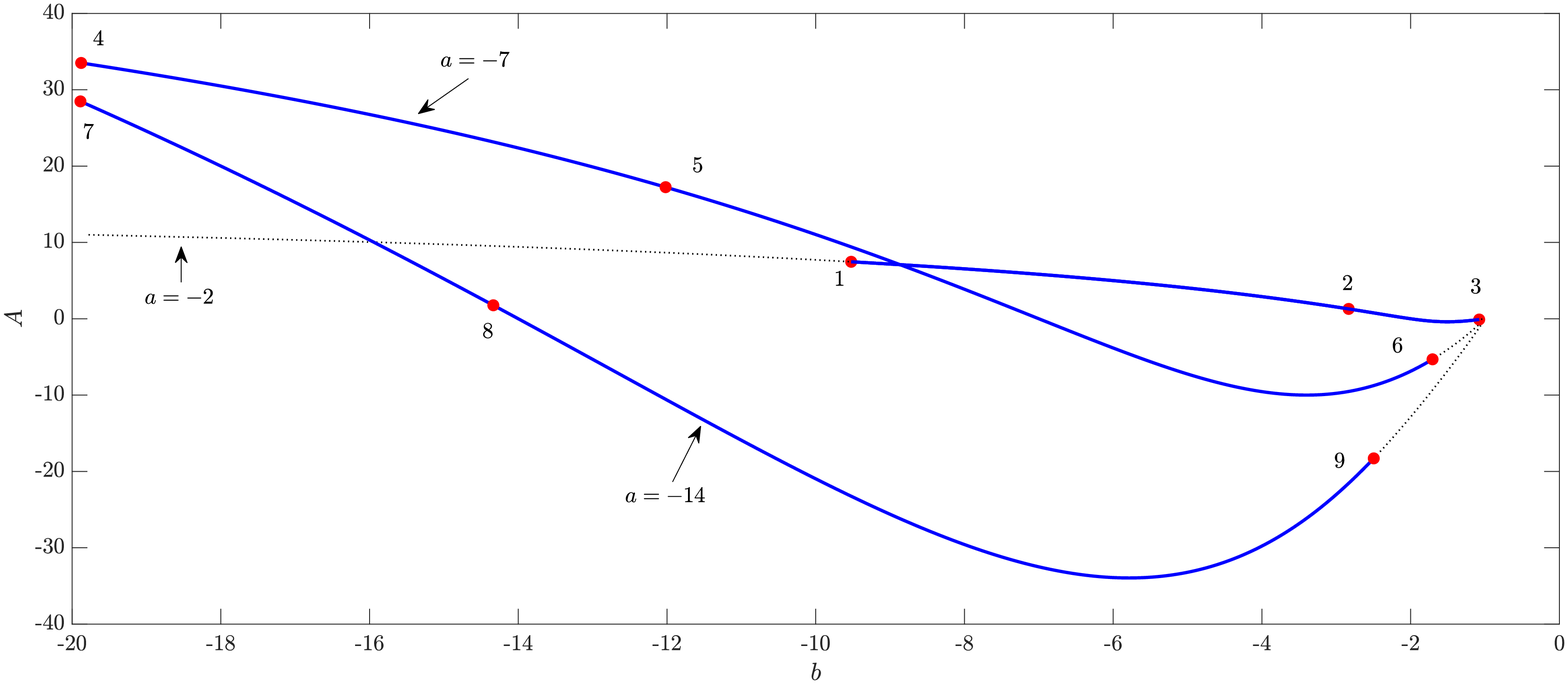}
  \includegraphics[width=\textwidth]{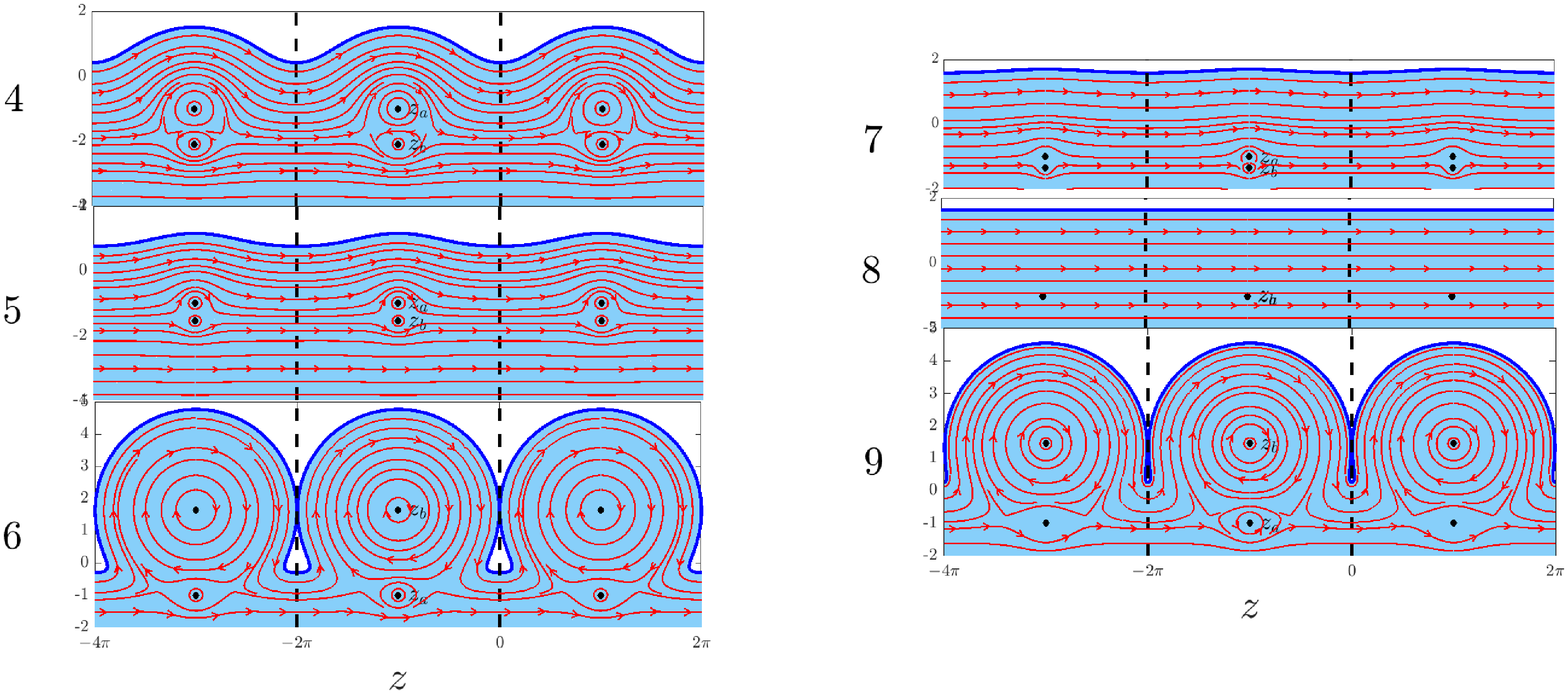}
  \caption{Solution structure for inline vortices, $a=-14,-7,-2$. The top panel shows $A$ as $b$ is varied. The solid lines indicate solutions of \eqref{conicA} that result in a univalent mapping. Thin dotted lines indicate solutions of \eqref{conicA} that do not result in valid solutions. The numbered profiles below the main panel correspond to the circular makers in the $(b,A)$ solution space.}
  \label{fig:complete_inline_bifurcation}
\end{figure}

\begin{figure}
  \centering
  \includegraphics[width=\textwidth]{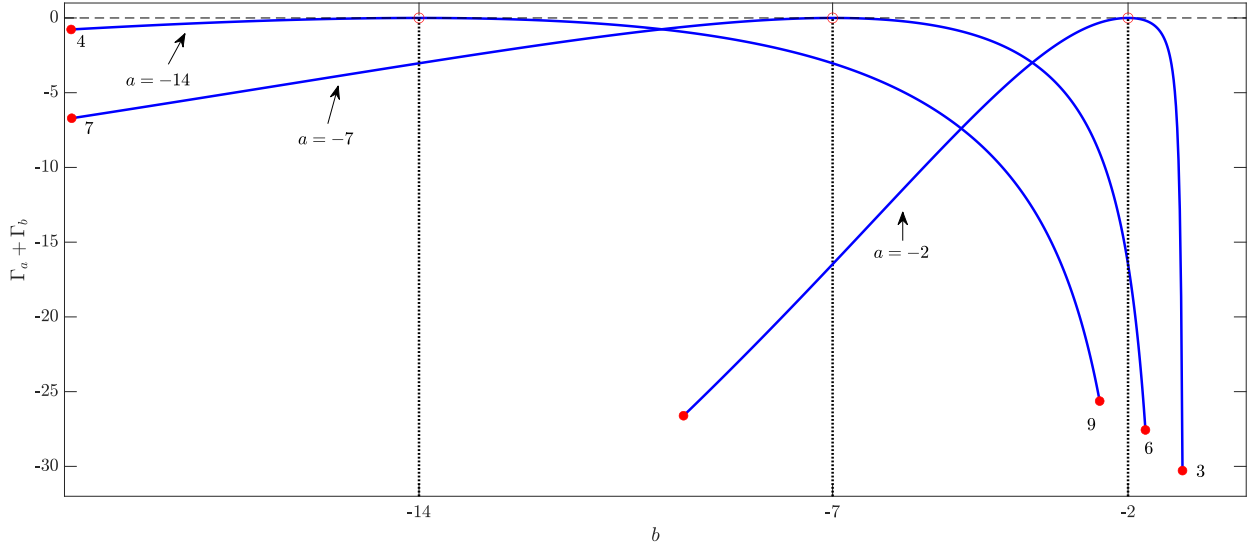}
  \caption{Total circulation of inline vortices, $a=-14,-7,-2$. The labels and solid markers indicate solutions in figure~\ref{fig:complete_inline_bifurcation}. The vertical dotted lines indicate where $a=b$ on each branch and the hollow circular markers indicate where there is no solution. The dashed horizontal line indicates the line $\Gamma_a + \Gamma_b = 0$.}
  \label{fig:complete_inline_bifurcation_circulation}
\end{figure}

For different values of $a$ we can vary $b$ and larger portions of the solution branch result in valid mappings. We emphasise that in each case only one root branch results in valid univalent solutions. Figure~\ref{fig:complete_inline_bifurcation} shows the valid branch for values of $a=-14,-7$ and -2. As can be seen in the main panel, the $a=-14,-7$ solution branches have a longer range of validity as $b\to-\infty$ but terminate at a lower value on the right side of the curve. In each case the limiting profiles, i.e. profiles 6,9 (and profile 3 in fig~\ref{fig:inline_bifurcation}) self-intersect with an adjacent periodic window. As $b\to\infty$ a solution persists for $a=-14,-7$, corresponding to an elevation profile. As $b\to a$, $A,B\to 0$ (see \S~\ref{sec:limit1}) and the profile becomes flat, as shown in profile 8 when $a=-14,b=-14.334$. This is because $A = B = 0$, and hence $\Gamma_a + \Gamma_b = 0$ as shown in figure~\ref{fig:complete_inline_bifurcation_circulation}.

\begin{figure}
  \centering
  \includegraphics[width=\textwidth]{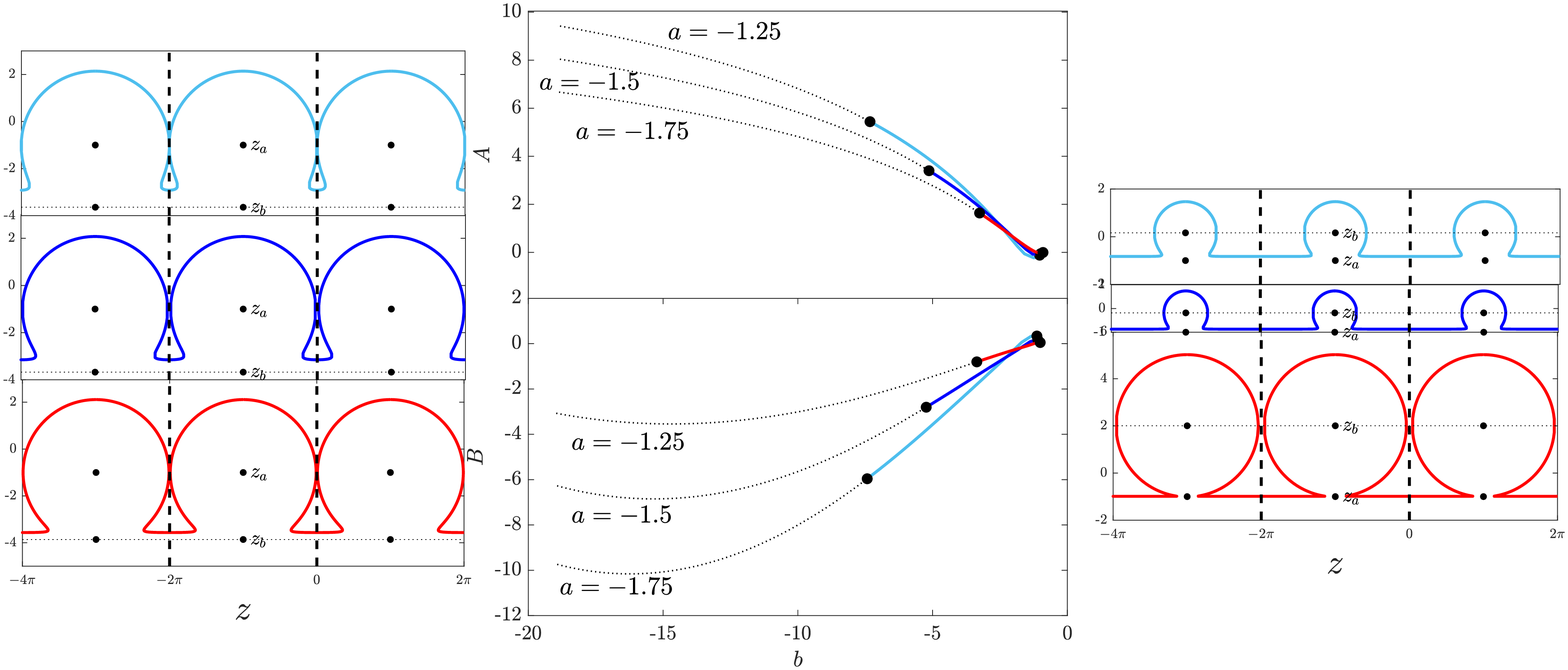}
  \caption{Solution structure of inline vortices, $a=-1.75,-1.5,-1.25$. The top middle panel shows $A$ and the bottom panel shows $B$, as $b$ is varied. The solid lines indicate solutions of \eqref{conicA} that give a univalent mapping. Thin dotted lines indicate solutions of \eqref{conicA} that do not result in valid solutions. The profiles on the left correspond to the markers furthest to the left in the main panel and the profiles on the right correspond to the right.}
  \label{fig:complete_inline_bifurcation_less_1}
\end{figure}

Exploring the solution space further, as $b\to 1$ it is found 
that the curves all collapse to $(A,B) = (0,0)$ 
as discussed in \S~\ref{sec:limit2}. 
However, this limit will depend on the value of $a$ as demonstrated in figure~\ref{fig:complete_inline_bifurcation_less_1}. 
Here we show the how the range of valid solutions 
in $b$, shrinks as $a\to -1$. The lower limit occurs 
at larger values of $b$ (profiles on the left in
 figure~\ref{fig:complete_inline_bifurcation_less_1})
 and the upper limit appears to increase towards $b=-1$ 
 (profiles on the right in figure~\ref{fig:complete_inline_bifurcation_less_1}). 
 The heights of the upper limit profiles do not change monotonically as $a$
  increases towards -1, so although it appears they are the curves are 
  converging on the same solution, they only converge in the limit as both $a,b\to -1$.

The limiting profiles are qualitatively different to those of \cite{crowdy2010steady}. For a single vortex, a cusp would appear in the middle of the periodic window for a critical value of $a$, beyond which a univalent mapping is not possible. This feature is not observed here. Instead, a
profile intersects with 
that in 
an adjacent period window.

%

\subsection{The staggered vortex street}
Staggered vortices require that $ab > 0$. As shown in figure~\ref{fig:rootregions} there are regions in $(a,b)$ space that result in a unique solution. 
This is explored further by examining the solution 
space for a fixed value of $a$ and then varying $b$, as done previously for the inline case.

\begin{figure}
  \centering
  \includegraphics[scale=0.4]{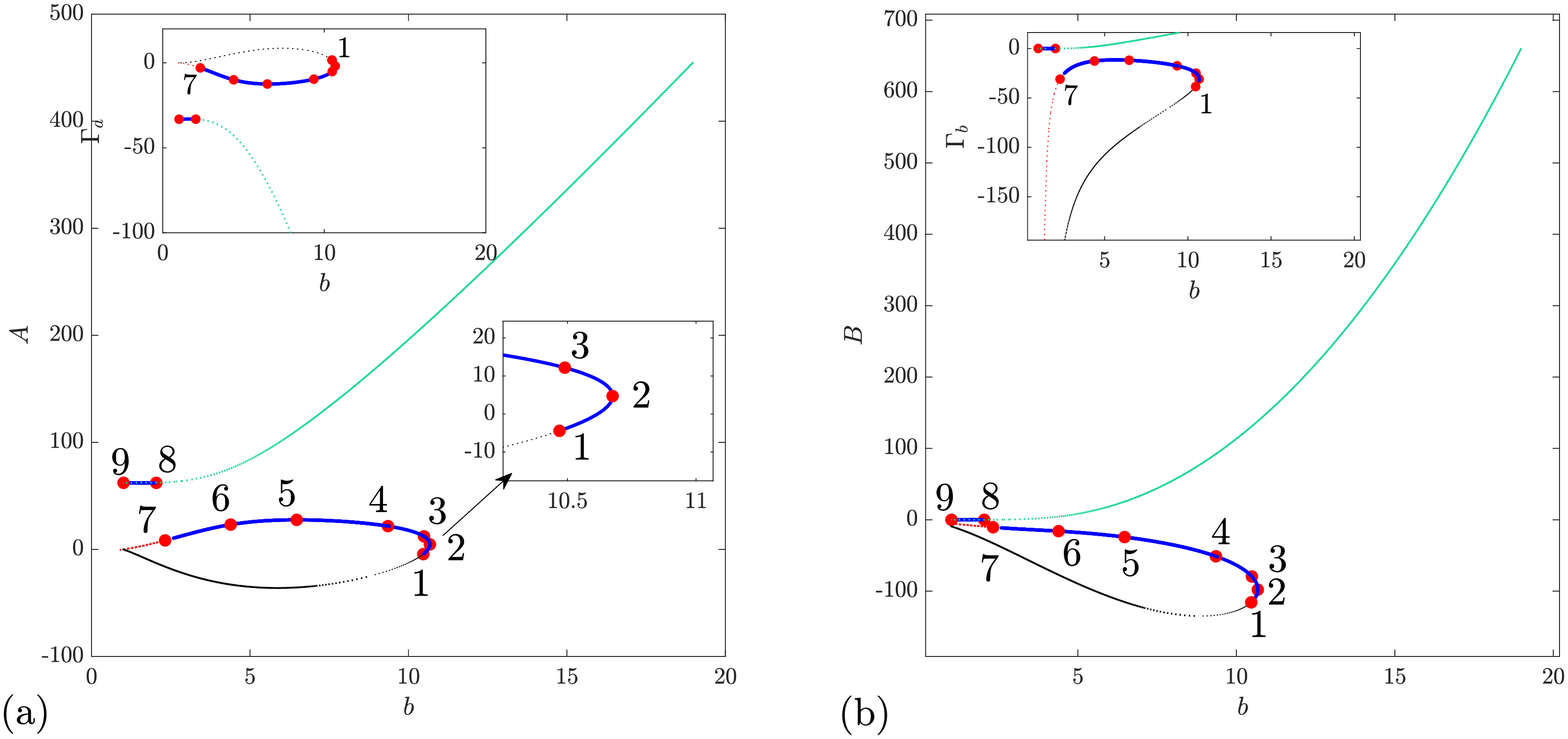}
  \includegraphics[scale=0.35]{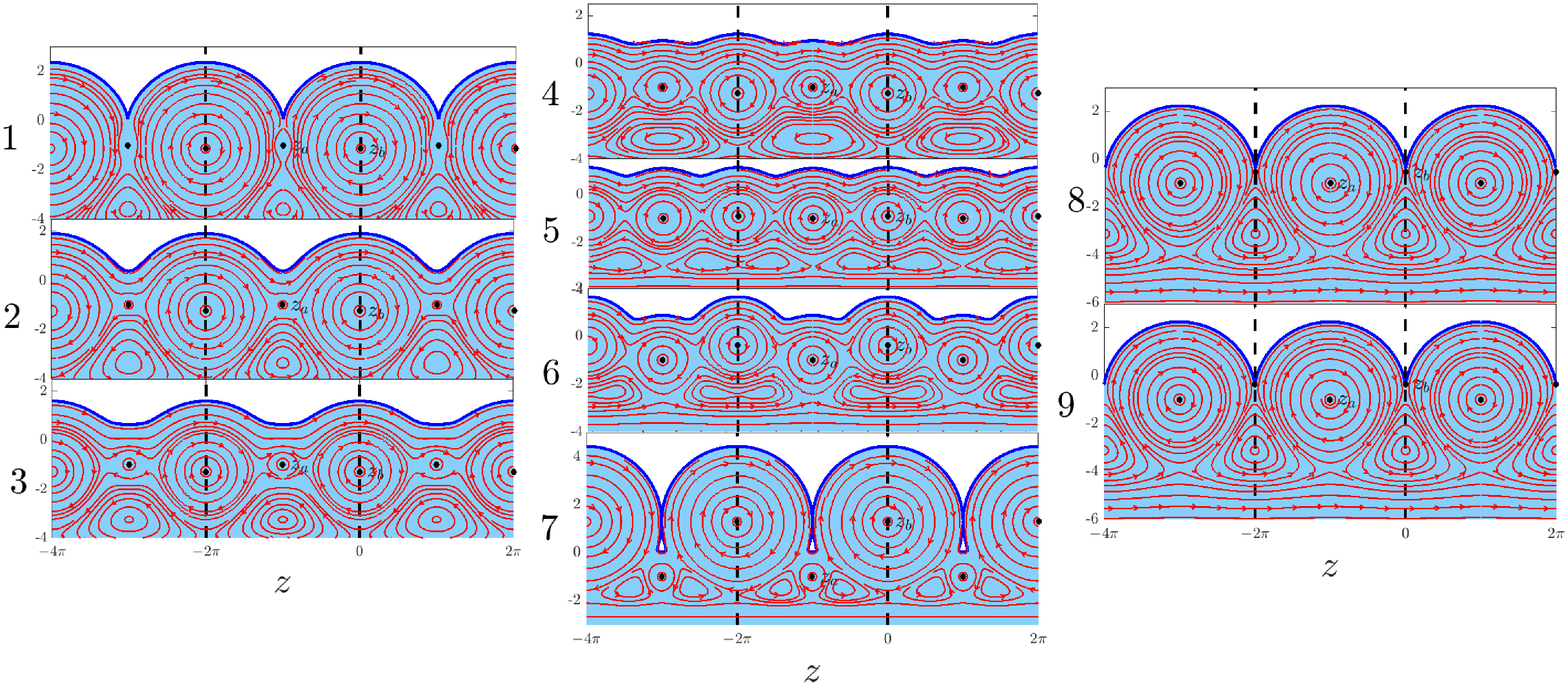}
  \caption{Solution structure for staggered vortices, $a=-7$. (a)/(b) shows plots of $A/B$, respectively, as $b$ is varied. The inset diagrams show the variation of the vortex strength, $\Gamma_a$/$\Gamma_b$ in panels (a)/(b) respectively. The solid lines indicate solutions of \eqref{conicA} that result in a univalent mapping. Thin dotted lines indicate solutions of \eqref{conicA} that do not result in a valid solution. The streamlines and profiles are show for the solutions labelled 1,2 and 3 in panel (a) as well as the positions of the vortices.}
  \label{fig:staggered_bifurcation}
\end{figure}

Figure~\ref{fig:staggered_bifurcation} shows the values of $A$ and $B$, panels (a) and (b) respectively, when $a=-7$. For sufficiently large $b$ there is only one single root of \eqref{conicA} which does not represent a valid solution. At $b\approx  10.68$ two additional solution branches appear via a fold bifurcation, both corresponding to valid solutions, as seen in the profiles labelled 1,2,3. The lower branch is only a valid solution until the profile develops a cusp in the middle of the period window, as seen in the profile labelled 1. This is similar to the limiting profiles in \cite{crowdy2010steady} and occur at parameter values $(a,b)$ such that $z'(1) = 0$.

  
Continuing on the upper branch as $b$ decreases, the profiles become bi-modal, with two distinct profile peaks, see profile 4, until $b=-a$, which corresponds to two horizontally aligned vortices, as seen in panel 5, where the profile is uni-modal. Decreasing $b$ further results in more distinct bi-modal wave profiles, see profile 6, until the branch reaches a termination point at $b\approx 2.3235$ when the interface self-intersects with an adjacent periodic window, see profile 7.

Interestingly, for values of $b$ less than this value there is a small portion of the other branch which allows valid solutions; see profiles 8 and 9. These profiles are similar in that both 
have a cusp at the edge of the periodic window, corresponding to $Z'(-1) = 0$. In profile 9, $z_b$ is close to the cusp, which is as expected as $b\to 1$, however as seen by the inset of panel (b), the strength of the vortex at $z_b$ in this limit is zero, rendering this vortex harmless.

\begin{figure}
  \centering
  \includegraphics[width=\textwidth]{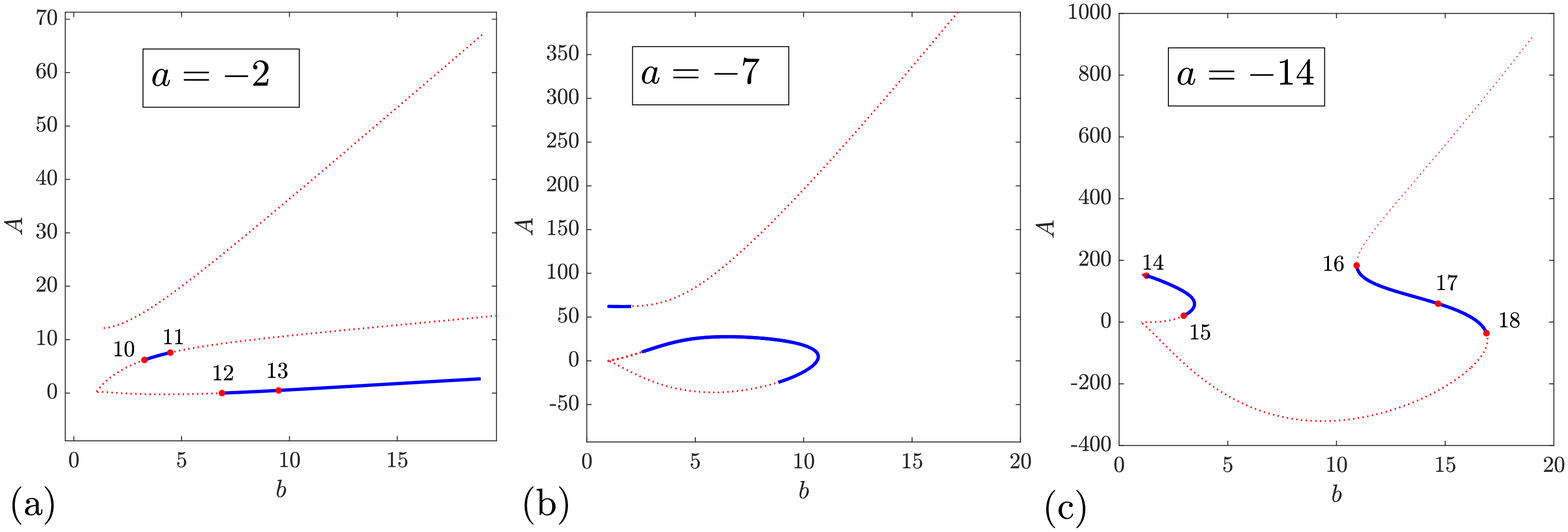}
  \includegraphics[width=\textwidth]{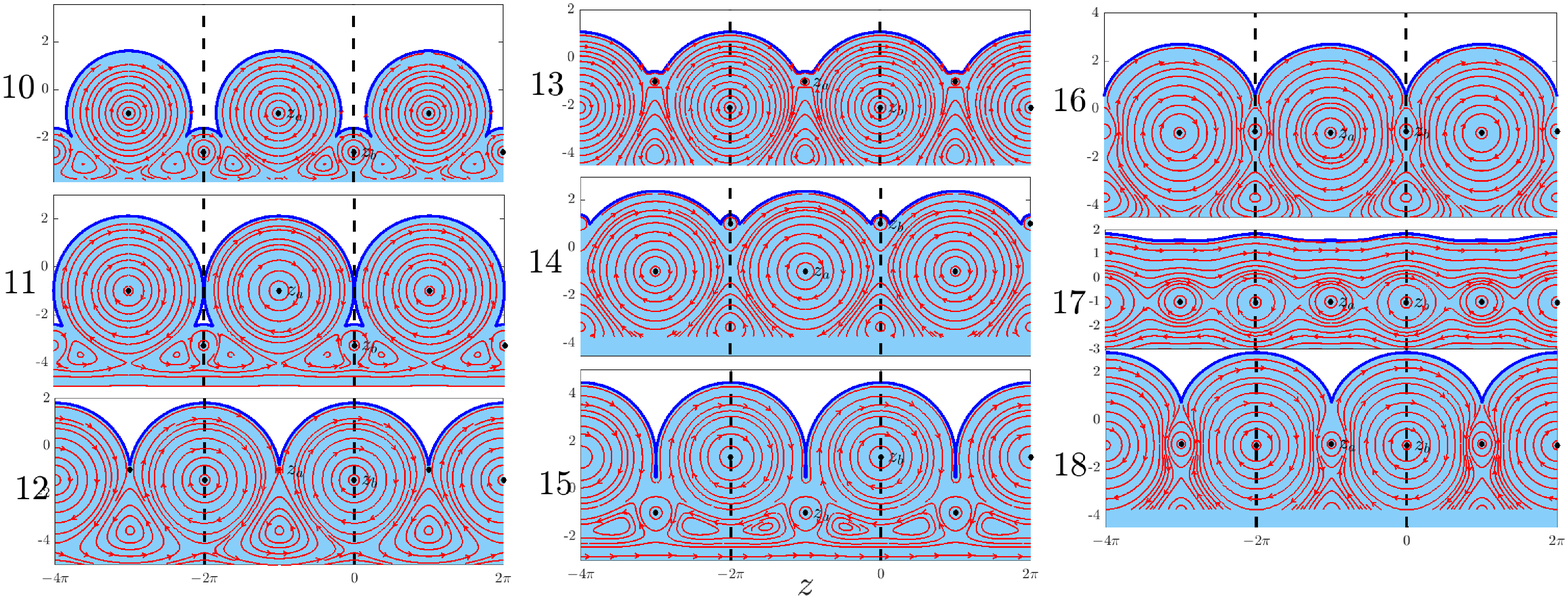}
  \caption{Solution structure for staggered vortices, (a) $a=-14$, (b) $a=-7$ and (c) $a=-2$. The solid lines indicate solutions of \eqref{conicA} that result in a univalent mapping. Thin dotted lines indicate solutions of \eqref{conicA} that do not result in a valid solution. The streamlines and profiles are shown for the solutions labelled 10-18.}
  \label{fig:complete_staggered_bifurcation}
\end{figure}

This bifurcation structure for $a=-7$ is not generic as $a$ is varied. Figure~\ref{fig:complete_staggered_bifurcation} shows the structure for $a=-14,-7,-2$, in panels (a),(b) and (c) respectively. When $a = -14$ there are always three roots and the fold bifurcation present when $a=-7$ ceases to exist. The profiles on the portion of the branches that are physically
admissible are significantly different to the $a=-7$ case.

The `upper' branch in panel (a) has no physically
admissible solutions, and the `middle' branch terminates on the `left' when the profile develops a cusp close to the edge of the periodic window, see panel 10, and terminates on the `right' when the interface intersects with the profile in an adjacent period window. 
The `lower' branch, as $b$ increases from 1, starts to produce a physically
admissible solution when a cusp develops in the middle of the period window, see panel 12 and for the parameter values we sampled continue to provide a physically
admissible solution as $b$ increases, resulting in a bi-modal wave profile, as seen in panel 13.

The structure changes again when $a$ becomes smaller as shown in panel (c), when $a=-2$. The different roots interact in a non-trivial manner through a number of different fold bifurcations. Starting at small $b>1$, profile 14 shows that the `upper' branch starts when there is a cusp near the edge of the period window, continues through a fold and eventually terminates when the profile self-intersects, see profile 15. There is a large region of $b$-values which does not produce a physically
admissible solution until a cusp develops at the edge of window, see profile 16, and then there is a single branch of solutions, eventually terminating when a cusp develops at the middle of the period window, see profile 18. Interestingly, these termination points appear to coincide close to fold bifurcations.

The structure of the equilibria for the staggered vortex system
is clearly rich and intriguing. Because the discriminant of \eqref{conicA} can change sign, the number of real solutions varies which results in a non-trivial interaction of the solution branches. This results in  quite exotic profiles, containing cusps and self-intersections.
These do not have 
direct counterparts in the case of a single cotravelling vortex row \citep{crowdy2010steady}. 

\begin{figure}
  \centering
  \includegraphics[width=\textwidth]{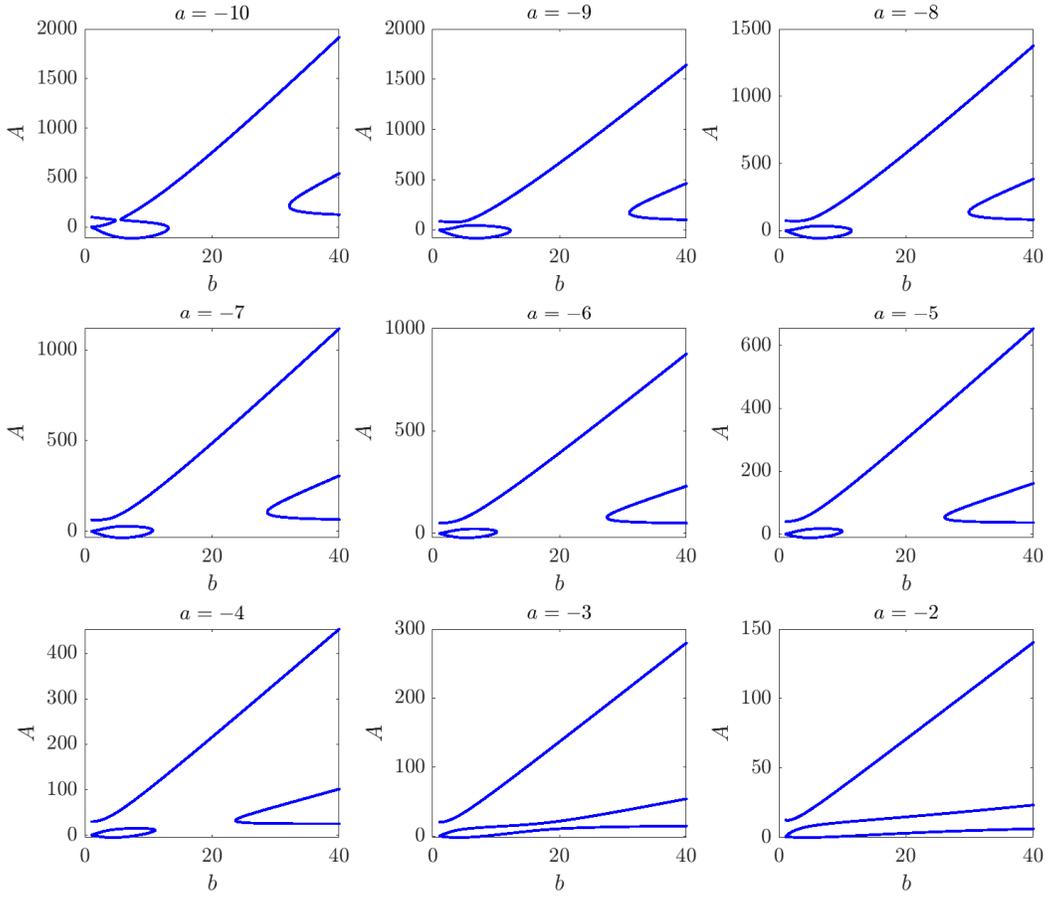}
  \caption{Solution structure for staggered vortices as $a$ varies from -10 to -2. In each panel $A$ is plotted against $b$ with the value of $a$ shown above. The roots are shown regardless of whether they represent valid physical mappings.}
  \label{fig:staggered_vary_a}
\end{figure}

\begin{figure}
  \centering
  \includegraphics[width=\textwidth]{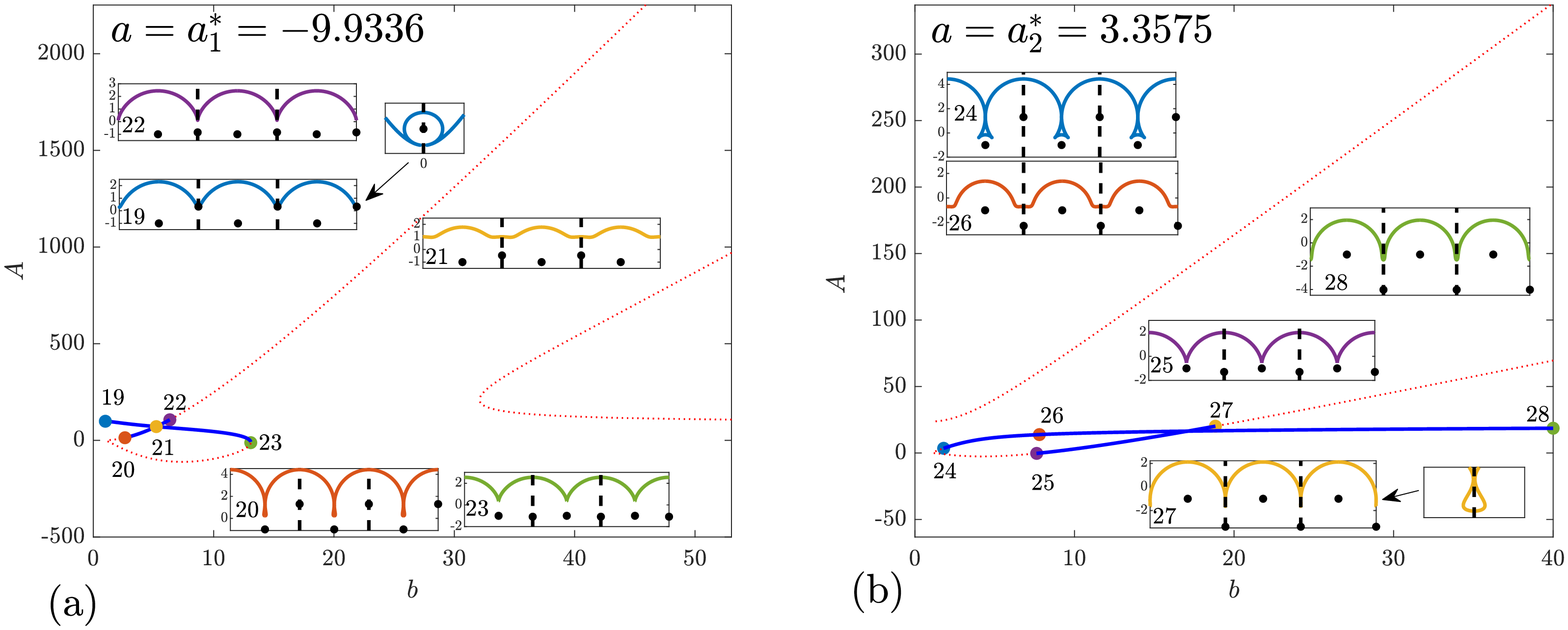} 
  \caption{Solution structure for $a=a_1^*,a_2^*$. Panels (a)/(b) show $A$ as $b$ is varied for $a = a_1^*$/$a_2^*$ respectively. The solid blue lines represent valid physical solutions. Limiting profiles are marked with labels.}
   \label{fig:a1star}
\end{figure}

Exploring this further, the parameter $a$ can be varied
to identify two transcritical bifurcations in the solution space. These occur when the two fold bifurcations collide and are
  shown in figure~\ref{fig:rootregions} as $(a_i^*,b_i^*)$. Figure~\ref{fig:staggered_vary_a} shows how the $(b,A)$ bifurcation diagram evolves as $a$ increases from -10 to -2. The first transcritical bifurcation occurs when the `hook' structure at small $b$ self-intersects and becomes a closed loop; see $a=-10$ and $a=-9$.  The first bifurcation occurs at $a_1^* = -9.9336$ to 4 d.p. The second transcritical bifurcation occurs when the `loop' structure intersects the lower branches for large $b$; see $a=-4$ and $a=-3$. This second bifurcation occurs at $a_2^* = -3.3575$ to 4.d.p. The bifurcation diagram for $a = a_1^*$ is shown in figure~\ref{fig:a1star}, where we identify sections of the curve that result in valid mappings (solid blue lines). The limiting profiles are indicated by the labels. 
  Particular attention is drawn to profile 19, which has a small unusual circular cusp at the edge of the periodic window. Finally,  no further bifurcations are observed as $a$ decreases from -14; the bifurcation structure remains robust as $a\to -\infty$.

\subsection{The limit $|a|,|b|\to \infty$}

            The limiting case
            where $|a|,|b|\to \infty$ is of significant interest.
            In this limit, 
             we expect the free surface to recede from the vortex rows into large positive $y$-values
             and become increasingly flat as deformation effects from the vortices weaken.
             It might be anticipated that the limiting equilibria would be 
            von K\'arm\'an vortex streets in unbounded shear; but it was 
            established in \S \ref{classic} that no such equilibria exist.
            
            Consider the case of staggered vortices with $b=-a$ with $a > 0$ and let $a \to \infty$.
            From the explicit expressions given in appendix \ref{appA} it can be shown that
            \begin{equation}
\lambda_2, \lambda_3 \sim {\cal O}(1/a), \qquad
\lambda_4 \sim 4 \log a - a^2, \qquad \lambda_5 \sim 4 \log a + a^2, \qquad \lambda_6 = -4 a^3 \log a
\end{equation}
so that the first of equations (\ref{circB}) becomes 
\begin{equation}
(4 \log a - a^2) A + (4 \log a + a^2)B -4 a^3 \log a \sim 0.
\end{equation}
From a similar analysis,
the second of equations (\ref{circB}) becomes
\begin{equation}
(4 \log a + a^2) A + (4 \log a - a^2)B + 4 a^3 \log a \sim 0.
\end{equation}
Together these two equations imply that 
\begin{equation}
A = - B \sim -2 a \log a.
\label{key10}
\end{equation}
Consequently, 
as $a \to \infty$,
\begin{equation}
\begin{split}
z= Z(\zeta) &\sim {\rm i} \left [ \log \zeta - {2a \log a \over \zeta-a} + {2 a \log a \over \zeta+a} \right ] + d 
\sim {\rm i} \left [ \log \zeta + {4  \log a}
 \right ] + d.
\end{split}
\label{key11}
\end{equation}
The condition
 $z_a = Z(1/a) = -{\rm i}$ then implies
\begin{equation}
-{\rm i} = 
Z(1/a) = {\rm i} \left [ \log(1/a)+ {Aa \over 1-a^2} + {B a \over 1-ab} \right ] + d
\end{equation}
implying that
\begin{equation}
\begin{split}
d 
\sim -{\rm i} (1+3 \log a).
\end{split}
\end{equation}
It follows from (\ref{key11}) that
\begin{equation}
 Z(\zeta)  \sim {\rm i} \left [ \log \zeta + {4  \log a}
 \right ] + d = {\rm i} \left [ \log \zeta + {4  \log a} - (1+3 \log a)
 \right ] = {\rm i} \log (\zeta a) - {\rm i}.
 \end{equation}
Suppose also that we insist that 
$z_b =Z(1/b)= \pi - {\rm i}(1+\lambda)$ then
\begin{equation}
{\rm i} \left [ \log(1/b) + {Ab \over 1-ab} + {Bb \over 1-b^2} -1- \log(1/a) - {Aa \over 1-a^2}
-{Ba \over 1-ab} \right ] = \pi -{\rm i}(1+\lambda).
\end{equation}
On use of (\ref{key10}) it follows from this that $\lambda \to 0$ so that
the two vortices  per period tend to $y=-1$ and are separated by distance $\pi$.
From (\ref{circA}) and (\ref{circB}),
\begin{equation}
-{{\rm i} \Gamma_a \over 2\pi} = -{A  Z'(1/a) \over 2a^2} \sim - {{\rm i} A  \over 2a}, \qquad
-{{\rm i} \Gamma_b \over 2\pi} = -{B  Z'(1/b) \over 2b^2} \sim - {{\rm i} B \over 2b},
\label{circB2}
\end{equation}
where the fact that, as $a \to \infty$, $Z(\zeta) \sim {\rm i} \log \zeta + {\rm constant}$ has been used.
But this means that
\begin{equation}
\Gamma_a \sim \Gamma_b = -2 \pi  \log a.
\end{equation}
The limiting configuration is not therefore the degenerate staggered
von K\'arm\'an vortex street in unbounded shear found in \S \ref{classic}. Rather, it is 
a single vortex row, with period $\pi$, of identical point vortices
with circulation $\Gamma_a$. It therefore falls within the class of solutions considered by \cite{Nelson};
indeed, it is easy to verify that (\ref{key10}) is consistent with equation (22) of \cite{Nelson} as an analogous 
parameter $a \to \infty$ in that study.
The new staggered equilibria found here can therefore be viewed as a steady ``pairing mode''
bifurcation from the latter solutions, that is, a class of subharmonic
bifurcations wherein adjacent pairs of vortices in a period-$\pi$ equilibrium found by \cite{Nelson}
displace separately to destroy the original periodicity forming instead 
 one of the $2\pi$-periodic
generalized equilibria found here.

An analysis of the inline case follows similarly. 
 In this case, for $b \approx a \to \infty$ the limiting configuration
is found to be an inline, or symmetric, von K\'arm\'an vortex street of vanishing aspect ratio $\lambda \to 0$ 
corresponding to the situation where the two vortex rows sit on top of each other and eventually
cancel each other out.

The significance of all these observations is that while the results of \S \ref{classic}
show that non-trivial equilibria generalizing the
classical von K\'arm\'an vortex streets do not exist in unbounded linear shear flow, a rich array of
steadily-travelling equilibria exists when a cotravelling free surface is also present.

            \section{Discussion}\label{sec:discussion}

There has been much recent interest in the problem of water waves with vorticity
\citep{MilesReview} and analytical solutions are rare.
This paper has unveiled
 a novel two-parameter family of analytical
 solutions for steadily-travelling water waves with uniform vorticity
 and superposed von K\'arm\'an point vortex streets.
 These solutions are direct extensions of earlier work on water waves with uniform
 vorticity and single cotravelling vortex row found by \cite{crowdy2010steady}.
 All these solutions fall within ``case 1'' of a 3-case categorization of water
 waves with vorticity recently set out by \cite{Recent}.

Fundamental to constructing the equilibria is finding the 
solution of a pair of algebraic nonlinear equations, which in the case of two vortices per period, has either one or three real roots that can be written and calculated in exact analytic form. 
The solution space for inline configurations is simpler in that there is always three real roots but only one root branch results in a univalent conformal map. These solution branches terminate when the wave profile intersects with that in an adjacent period window. 
For staggered configurations, the solution space is more complicated as there are regions of parameter space where only a single real root exists. For these configurations limiting profiles can exist where cusps form, or where the interface intersects, resulting in a rich variety of solutions, including bi-modal wave profiles and fold bifurcations.


The stability of the various equilibria found here is clearly of great interest,
but requires detailed investigation and has not been studied here.
 \cite{CrowdyClokeStab} have studied the linear stability of analogous case 1 solutions
 in the radial (vortex) geometry and found that exact solutions within this class can be
 linearly stable. 
 The methods used in that study are easily adaptable to study the stability of the new water
 wave equilibria found here. 
 More
 recently, \cite{blyth2022stability} have
 calculated the exact linear stability spectrum of the waves described in \cite{hur2020exact}  -- which fit into case 3 of the taxonomy of \cite{Recent} -- 
 and those methods should also be generalizable to the solutions presented here.

The method we describe here can easily be extended to $n$ submerged point vortex rows. 
For $n$-vortices per period there will arise an $n$-parameter family of solutions with a system of 
$n$ quadratic relations to be solved. From B\'{e}zout's theorem, this means that there will be potentially $2^n$ possible solutions; the complexity of the solution space exponentially increases as $n$ increases. For $n=3$, the equations are easy to solve using a symbolic algebra package but the complexity of the closed-form solution means that the computer time taken to solve the system symbolically, and then convert to double precision numbers means that it is inefficient to calculate the roots in this way. For example, choosing $(a,b,c) = (-2,-3,-4)$ where $c$ represents the parameter of the third vortex take ~200 computer seconds to compute the roots symbolically and then convert to a double precision number. Interestingly, 
by an extension of the degeneracies evident here, and in the earlier study of \cite{Nelson},
 5 roots are obtained, rather than the 8 predicted.
So far, after limited investigation,
we have observed that none of these roots result in a valid univalent map.
However, we
 leave the existence of steady $n$-vortex configurations as an open question.
 
 More broadly, it is worth mentioning that other extensions of the classical 
 von K\'arm\'an point vortex streets have been found. \cite{VK} have found 
 analytical solutions for steadily-travelling
  streets of so-called hollow vortices which are finite-area regions of constant pressure having
  non-zero circulation around them. These solutions can be understood
  as regularized  von K\'arm\'an vortex streets where the singular point vortices
  are replaced by finite-area vortices for which the associated velocity fields are everywhere finite.
 This hollow vortex model has much in common with the water wave problem in that the boundary condition
 on the boundary of a hollow vortex, which also neighbours a constant-pressure region,
  is akin to that on the free surface between a water wave
  and a constant-pressure region.
  \cite{Roenby} discuss the similarities between these two problems.
   In view of the new equilibrium solutions found here, and the new case-2 solutions
  for submerged von K\'arm\'an point vortex streets beneath a free surface
  recently found in \cite{Recent}, it is of interest
  to examine if the hollow vortex street equilibria of  \cite{VK} can be generalized to incorporate
  steady translation beneath a cotravelling free surface wave in the spirit of the present study.
  Such matters await further investigation.
  
  The effects of gravity and surface tension have been ignored here, 
  but how they will alter the new wave solutions found here is clearly of interest.
  It should be possible
to add the effect
of weak gravity as a regular perturbation, an analysis that should be greatly facilitated by 
having available closed-form expressions for the leading-order equilibria.
Such analyses have recently been carried out for constant-vorticity leading-order solutions
by  \cite{MilesNEW}; see also \cite{Ambrose} who added weak gravity to irrotational capillary
waves.
Asymptotic analyses of the effects of weak capillarity \citep{Chapman}
on the solutions here will similarly be made easier
by the closed-form description of the equilibria.


\appendix

\section{Coefficients of \eqref{conicA} for $A$ and $B$} \label{appA}
The coefficients of \eqref{conicA} are found to be
\begin{multline}
  \lambda_1 = 0,\:\lambda_2 = \frac{2a^4(a^2 - 1)}{(a - b)(ab - 1)^3},\: \lambda_3 = \frac{2a^4(b^2 - 1)}{(a - b)(ab - 1)^3}, \lambda_4 = \frac{a^2(2a^2\log(a^2) - a^4 + 1)}{(a^2 - 1)^2},\\ \lambda_5 = \frac{2a^3(ab + a^2\log(a^2) - a^3b + a^2 - ab\log(a^2) - 1)}{(a - b)(ab - 1)^2},\:\lambda_6 = -2a^3\log(a^2)
\end{multline}
and 
\begin{multline}
  \mu_2 = \frac{2b^4(b^2 - 1)}{(b - a)(ab - 1)^3},\: \mu_2 = 0,\: \mu_3 = \frac{2b^4(a^2 - 1)}{(b - a)(ab - 1)^3},\\\mu_4= \frac{2b^3(ab + b^2\log(b^2) - b^3a + b^2 - ab\log(b^2) - 1)}{(b - a)(ab - 1)^2},\:\mu_5 = \frac{b^2(2b^2\log(b^2) - b^4 + 1)}{(b^2 - 1)^2},\\ \mu_6 = -2b^3\log(b^2)
\end{multline}

\vskip2pc                                                                                                
\bibliographystyle{jfm}
\bibliography{bibref}

\end{document}